\documentclass[usenatbib]{mnras}
\usepackage{graphicx}% Include figure files
\usepackage{dcolumn}% Align table columns on decimal point
\usepackage{amssymb} %maths
\usepackage{amsmath}
\usepackage{ctable}
\usepackage{float}
\usepackage{bm}% bold math
\usepackage{float}
\usepackage{epsfig}
\usepackage{epstopdf}
\usepackage{epsf,color}
\usepackage{comment}
\usepackage{aas_macros}
\usepackage{url}
\usepackage{soul}
\usepackage{color, colortbl}
\usepackage{widetext}
\usepackage[T1]{fontenc}
\newcommand{\cmnt}[1]{}

\makeatletter % Workaround to only color the year for cite commands
  \patchcmd{\NAT@citex}
    {\@citea\NAT@hyper@{%
      \NAT@nmfmt{\NAT@nm}%
      \hyper@natlinkbreak{\NAT@aysep\NAT@spacechar}{\@citeb\@extra@b@citeb}%
      \NAT@date}}
    {\@citea\NAT@nmfmt{\NAT@nm}%
    \NAT@aysep\NAT@spacechar\NAT@hyper@{\NAT@date}}{}{}

  \patchcmd{\NAT@citex}
    {\@citea\NAT@hyper@{%
      \NAT@nmfmt{\NAT@nm}%
      \hyper@natlinkbreak{\NAT@spacechar\NAT@@open\if*#1*\else#1\NAT@spacechar\fi}%
        {\@citeb\@extra@b@citeb}%
      \NAT@date}}
    {\@citea\NAT@nmfmt{\NAT@nm}%
    \NAT@spacechar\NAT@@open\if*#1*\else#1\NAT@spacechar\fi\NAT@hyper@{\NAT@date}}
    {}{}
\makeatother

\usepackage[normalem]{ulem}
\usepackage{color}

\usepackage{tgtermes}
\newcommand{\hii}{\ion{H}{ii}}
\newcommand{\hi}{\ion{H}{i}}
\newcommand{\hei}{\ion{He}{i}}
\newcommand{\heii}{\ion{He}{ii}}
\newcommand{\heiii}{\ion{He}{iii}}

\newcommand{\oi}{\ion{O}{i}}
\newcommand{\oiii}{\ion{O}{iii}}
\newcommand{\oii}{\ion{O}{ii}}

\newcommand{\nii}{\ion{N}{ii}}
\newcommand{\siii}{\ion{S}{iii}}
\newcommand{\angstrom}{\mbox{\AA}}

\title[Ionized ISM emission lines]{Efficient simulations of ionized ISM emission lines: A detailed comparison between the FIRE high-redshift suite and observations}

\author[S. Yang et al.]{%
Shengqi Yang,$^{1}$\thanks{E-mail: \href{mailto:syang@carnegiescience.edu}{syang@carnegiescience.edu}}
Adam Lidz,$^{2}$
Aaron Smith,$^{3}$
Andrew Benson$^{1}$
and
Hui Li$^{4,5}$
\\
$^{1}$Carnegie Observatories, 813 Santa Barbara Street, Pasadena, CA 91101, USA\\
$^{2}$Department of Physics and Astronomy, University of Pennsylvania, 209 South 33rd Street, Philadelphia, PA 19104, USA\\
$^{3}$Center for Astrophysics $\vert$ Harvard $\&$ Smithsonian, 60 Garden Street, Cambridge, MA 02138, USA\\
$^{4}$Department of Astronomy, Columbia University, New York, NY 10027, USA\\
$^{5}$Department of Astronomy, Tsinghua University, Beijing 100084, China\\
}

\date{Accepted XXX. Received YYY; in original form ZZZ}

\pubyear{2023}

\begin{document}
\label{firstpage}
\pagerange{\pageref{firstpage}--\pageref{lastpage}}
\maketitle

\begin{abstract}
The Atacama Large Millimeter/Submillimeter Array (ALMA) in the sub-millimeter and the \textit{James Webb Space Telescope} (\textit{JWST}) in the infrared have achieved robust spectroscopic detections of emission lines from the interstellar medium (ISM) in some of the first galaxies. These unprecedented measurements provide valuable information regarding the ISM properties, stellar populations, galaxy morphologies, and kinematics in these high-redshift galaxies and, in principle, offer powerful tests of state-of-the-art galaxy formation models, as implemented in hydrodynamical simulations.   
To facilitate direct comparisons between simulations and observations, we develop a fast post-processing pipeline to predict line emission from the \hii\ regions around simulated star particles, accounting for spatial variations in the surrounding gas density, metallicity, and incident radiation spectrum. 
 Our ISM line emission model currently captures H$\alpha$, H$\beta$, and all of the [\oiii] and [\oii] lines targeted by ALMA and \textit{JWST} at $z>6$.  We illustrate the power of this approach by applying our line emission model to the publicly available Feedback In Realistic Environments (FIRE) high-$z$ simulation suite and perform a detailed comparison with current observations. We show that the FIRE mass--metallicity relation is in $1\sigma$ agreement with ALMA/\textit{JWST} measurements after accounting for the inhomogeneities in the ISM properties. We also quantitatively validate the description of the one-zone model, which is widely used for interpreting
 [\oiii] and H$\beta$ line luminosity measurements.
 This model is publicly available and can be implemented on top of a broad range of galaxy formation simulations for comparison with \textit{JWST} and ALMA measurements.
\end{abstract}


%%%%%%%%%%%%%%%%%%%%%%%%%%%%%%%%%%%%%%%%%%%%%%%%%%

%%%%%%%%%%%%%%%%% BODY OF PAPER %%%%%%%%%%%%%%%%%%

\section{Introduction}
A crucial step for understanding the reionization process and the subsequent formation of large-scale structure of the Universe, is to make direct observations of the first galaxies. The Hubble Space Telescope (HST) has identified $\sim$ 1000 photometric candidate $z > 6$ galaxies, and measured their abundance versus UV luminosity and cosmic time/redshift \citep[e.g.][]{2015ApJ...803...34B,2015ApJ...810...71F,2017ApJ...835..113L,2017ApJ...843..129B,2018MNRAS.479.5184A,2018ApJ...855..105O,2018ApJ...854...73I,2019MNRAS.486.3805B,2022ApJ...940...55B}.
With new facilities such as ALMA and \textit{JWST}, we can move beyond simply counting these galaxies and measure comprehensive emission line spectra. Currently, ALMA has measured submillimeter lines from the ISM in tens of $z \sim 6$--9 galaxies, providing valuable spectroscopic confirmations of photometric candidates \citep[e.g.][and references therein]{2016Sci...352.1559I,2017ApJ...837L..21L,2018Natur.557..392H,2019MNRAS.487L..81L,2019PASJ...71...71H,2020ApJ...896...93H,2022MNRAS.515.1751W}. New \textit{JWST} data have yielded comprehensive spectroscopic measurements of rest-frame optical emission lines from the ISM in multiple $z \sim 8$ galaxies, with larger galaxy samples forthcoming in the near future \citep{ArellanoCordova2022,2022arXiv220708778C,2022arXiv220712375C,2022arXiv220803281T,2022arXiv220713693K,2022arXiv220713020R,2022arXiv220710034S,2022arXiv220712388T,2022arXiv220714265T,2022arXiv221202890H}. The [\oiii] submillimeter lines resolved by ALMA and the [\oiii] rest-frame optical lines, [\oii] 3727,\,29\,$\angstrom$ doublet, H$\alpha$, and H$\beta$ lines probed by \textit{JWST} are sensitive probes of the properties of interstellar gas in these early galaxies. For example, the luminosity ratio between the [\oiii] 88\,$\mu$m and 52\,$\mu$m lines is a good diagnostic of the density of the \hii region. Furthermore, the luminosity ratios between the [\oiii] 5008\,$\angstrom$ and 88\,$\mu$m or the [\oiii] 4364\,$\angstrom$ lines are sensitive to the gas temperature. In addition, the ratio between the [\oiii] and H$\alpha$ or H$\beta$ line luminosities can be used to constrain the gas-phase metallicity \citep{2011piim.book.....D}. Moreover, the luminosity ratios between [\oiii] and [\oii] lines depend on the hardness of the radiation spectrum and may also correlate with the hydrogen ionizing photon escape fraction, $f_\mathrm{esc}$, which is an important quantity for understanding the reionization process, yet still highly uncertain \citep[e.g.][]{2016ApJ...829...99F,2018MNRAS.478.4851I,2020ApJ...902L..39B,2022MNRAS.510.4582N}. The current and upcoming measurements of multiple ISM emission lines from high-redshift galaxies therefore
provide valuable information and strong tests of galaxy formation models, including those from state-of-the-art hydrodynamic simulations.\par
Significant progress has been achieved in assigning line luminosities to numerical simulations of individual galaxies, capable of partially resolving the ISM in the simulated galaxies \citep[e.g.][]{2018MNRAS.481L..84M,2022MNRAS.514.3857K,Garaldi_2022,Smith_2022,2023MNRAS.tmpL...2K,2022arXiv221202522H,2023arXiv230102416N}. A widely adopted ISM line emission post-processing pipeline is to compute line luminosities on a grid of ISM parameters using a spectral synthesis code such as \textsc{Cloudy} \citep{2017RMxAA..53..385F} or \textsc{Mappings} \citep{1985ApJ...297..476B,1993ApJS...88..253S,2018ascl.soft07005S}, and to then assign line emission signals to each gas or star particle in the simulated galaxies through interpolation. Due to the large number of degrees-of-freedom involved in modeling ISM line emission, it is computationally expensive to create line signal lookup tables that cover all of the parameters of interest. More specifically, the incident stellar radiation spectrum is generally characterized by three parameters: The age, metallicity, and mass of the stellar particles. The important ISM properties for modeling line emission signals include gas density, metallicity, and temperature. In this case, the stellar population/ISM parameter space is six-dimensional, and so including ten grid zones per parameter requires evaluating $10^6$ models, which takes about 3500 CPU hours using \textsc{Cloudy}. 
The computing time increases exponentially with finer parameter space sampling, and it becomes difficult to explore still higher-dimensional parameter spaces, as may be necessary in some applications. Studies of ISM line emission from simulations are therefore forced to adopt oversimplified assumptions about the gas properties and the stellar radiation spectral shape in order to guarantee manageable lookup table calculations. They hence fail to take full advantage of the ISM and stellar population information provided by the simulations. The simplifications made in post-processing the line emission signals from simulations may also prohibit more robust and direct comparisons with observations. Recent efforts have also been made to model photoionization and reprocessed line emission throughout entire galaxies down to the resolution limit of hydrodynamical simulations \citep[e.g.][]{Katz2022,Smith2022COLT,Tacchella2022COLT}. The main advantage of this approach is the more self-consistent non-local radiation field and inclusion of processes such as non-equilibrium thermochemistry and dust extinction within three-dimensional geometries. However, galaxy-scale simulations are far from fully resolving compact \hii\ regions and radiative cooling scales. Therefore, sub-resolution density, temperature, dust, and ionization state structural information is subject to the limits of the state-of-the-art and certainly affect aspects of line emission predictions.
\par

In this work, we address the above challenges by introducing an analytical ISM emission line model, covering all of the [\oiii], [\oii], H$\alpha$, and H$\beta$ lines from the Epoch of Reionization (EoR) targeted by ALMA and \textit{JWST}. The advantage of this new ISM emission line model compared with other spectral synthesis codes is its high computational efficiency, allowing it to predict the line luminosity across huge numbers of simulation cells in a fast and self-consistent way. Our stripped-down approach focuses solely on a few important lines of interest and models the line emission from first principles. This methodology can help isolate and elucidate the most important physics involved, which can sometimes be obscured by more complex codes, such as \textsc{Cloudy}. Moreover, our flexible approach can be applied across many different ISM sub-grid models, which allow their application even to large volume simulations. It is therefore a necessary tool to facilitate direct ISM simulation--observation comparisons. As a first application of our model we use it to calculate ISM line emission from the primary galaxies in the publicly available FIRE high-$z$ simulation suite \citep{2016MNRAS.456.2140M,2018MNRAS.478.1694M,2022arXiv220206969W}. We model the \hii\ regions around each star particle in the simulation and treat each \hii region as an individual line emitter, with the ISM properties estimated from neighbouring gas particles. The spectrum of each stellar particle is calculated using the Flexible Stellar Population Synthesis code \cite[FSPS;][]{2009ApJ...699..486C,2010ApJ...712..833C}, which takes the stellar mass, stellar metallicity, and birth time directly from the FIRE zoom-in simulation as input. We then present detailed comparisons between predictions of the emission in multiple lines from the simulated FIRE galaxies and \textit{JWST}/ALMA observations. We also explore how inhomogeneities in the ISM properties, which are ignored in many previous works, can influence the interpretation of line emission measurements. We have made a code that implements our modeling publicly available at \url{https://github.com/Sheng-Qi-Yang/HIILines}. The code may be readily applied to other high-resolution simulations of galaxy formation.\par
The plan of this paper is as follows. In Section~\ref{sec:model}, we introduce our new ISM line emission model for [\oiii], [\oii], H$\alpha$, and H$\beta$ lines. In Section ~\ref{sec:FIRE}, we briefly introduce the FIRE high-$z$ suite and our method of assigning HII region properties to each stellar particle. We then combine the ISM emission line model with the FIRE simulations and perform detailed comparisons with \textit{JWST}/ALMA observations. In Section~\ref{sec:inhomo}, we study how inhomogeneities in the ISM properties may impact inferences
of these properties from line emission observations. We summarize the main results and discuss the caveats of our approach in Section~\ref{sec:discussion}.

\section{model}\label{sec:model}

The ionized ISM emission line model introduced in this work contains a 5-level description for the [\oiii] and [\oii] ions, refining the 3-level [\oiii] ion treatment in \cite{2020MNRAS.499.3417Y}. We refer the readers to \cite{2020MNRAS.499.3417Y} for a detailed discussion of the simpler version of the model employed there.\par
Throughout this work, we will use $n_{X}$ to denote the number density of particle $X$, and $n^X_i$ for the number density of particle $X$ in its $i^\mathrm{th}$ energy state, while $h\nu_{X}$ is the ionization energy to produce ion $X$. We will use $\nu^X_{ij}$ to denote the rest-frame frequency (in a vacuum) for the line emitted when ion $X$ decays from level $i$ to level $j$. The quantities $A^X_{ij}$ and $k^X_{ij}$ denote the spontaneous decay rate and the collisional (de-)excitation rate between energy levels $i$ and $j$ for particle $X$. Finally, $Q_\mathrm{X}=\int_{\nu_\mathrm{X}}^\infty L_\nu/(h\nu)d\nu$ gives the rate of generating ionizing photons (which ionize particle $X$), in terms of the incident spectrum with specific luminosity $L_\nu$. \par 
We employ the photo-ionization cross section $\sigma_\mathrm{X}(\nu)$ fitting formulas for particle $X$ in \cite{1996ApJ...465..487V}; case A and B recombination rate of \hii\ and \heii\  ($\alpha_\mathrm{B,\hii}$, $\alpha_\mathrm{A,\heii}$, and $\alpha_\mathrm{B,\heii}$), the rate of \heii\ recombination directly to the ground state ($\alpha_\mathrm{1,\heii}$), and the rate coefficients
for recombinations that result in H$\alpha$ and H$\beta$ line emissions ($\alpha_\mathrm{B,H\alpha}$, $\alpha_\mathrm{B,H\beta}$) from \cite{2011piim.book.....D}; charge transfer rates $\delta_\mathrm{\oii}'$ for \oiii$+$\hi$\leftrightarrow$\oii$+$\hii\ from \cite{2006agna.book.....O}; charge transfer rates $k_0$, $k_1$, $k_2$ for \hi$+$\oii$\rightarrow$\hii$+$\oi($^3P_2$), \hi$+$\oii$\rightarrow$\hii$+$\oi($^3P_1$), \hi$+$\oii$\rightarrow$\hii$+$\oi($^3P_0$), as well as the rate for the reverse reaction \hii$+$\oi($^3P_2$)$\rightarrow$\hi$+$\oii\ ($k_{0r}$) from \cite{2006agna.book.....O}. Information regarding \oiii\ and \oii\ ion energy levels in their ground electron configurations, as well as relevant radiative transfer rates, are adopted from \cite{2006agna.book.....O} and \cite{2011piim.book.....D}.\par
\subsection{\hii, \oiii, and \oii\ region volumes}\label{subsec:volume}
Consider a simple picture where a stellar population ionizes its nearby ISM, resulting in a spherically symmetric \hii\ region. Assuming the ISM is uniform in hydrogen number density $n_\mathrm{H}$, helium number density $n_\mathrm{He}$, and temperature $T$, the hydrogen and helium ionization--recombination equilibrium\footnote{Assuming ionization--recombination equilibrium among H and He is a good approximation because the recombination time-scales, $1/(\alpha_\mathrm{B,\hii}n_e)\approx1.2\times10^3$\,yr, $1/(\alpha_\mathrm{B,\heii}n_e)\approx1.2\times10^3$\,yr ($n_e=100$\,cm$^{-3}$, $T=10^4$\,K) is much shorter than the lifetime of O-stars.} can be expressed as:
    \begin{equation}\label{eq:HHebalance}
    \begin{split}
    \dfrac{n_\mathrm{\hi}}{4\pi r^2}\int_{\nu_\mathrm{\hi}}^\infty\dfrac{L_\nu}{h\nu}e^{-\tau_\nu}\sigma_\mathrm{\hi}(\nu)d\nu+&yn_\mathrm{\heii}n_e\alpha_\mathrm{1,\heii}\\
    +pn_\mathrm{\heii}n_e\alpha_\mathrm{B,\heii}&=n_\mathrm{\hii}n_e\alpha_\mathrm{B,\hii}(T)\,,\\
    \dfrac{n_\mathrm{\hei}}{4\pi r^2}\int_{\nu_\mathrm{\hei}}^\infty\dfrac{L_\nu}{h\nu}e^\mathrm{-\tau_\nu}\sigma_\mathrm{\hei}(\nu)d\nu+&(1-y)n_\mathrm{\heii}n_e\alpha_\mathrm{1,\heii}\\
    &=n_\mathrm{\heii}n_e\alpha_\mathrm{A,\heii}(T)\,.
    \end{split}
\end{equation}
Here $n_\mathrm{\hi}$ and $n_\mathrm{\hii}$ satisfies $n_\mathrm{\hi}+n_\mathrm{\hii}=n_\mathrm{H}$. We neglect the presence of doubly ionized helium for simplicity and assume $n_\mathrm{HeI}+n_\mathrm{HeII}=n_\mathrm{He}$, with the helium abundance characterized by the gas-phase metallicity $n_\mathrm{He}=(0.0737+0.0293Z)n_\mathrm{H}$ \citep{2004ApJS..153...75G}. Very hard incident spectra with photon energies higher than 54.42 eV can generate \heiii, but \heiii\ regions generally occupy only a tiny fractional volume of the entire \hii\ region, and therefore they do not have a significant influence on the lines that we model in this work. The number density of free electrons is therefore $n_e=n_\mathrm{\hii}+n_\mathrm{\heii}$.
\begin{equation}
    y\approx\dfrac{n_\mathrm{\hi}\sigma_\mathrm{\hi}(24.59\mathrm{eV}+k_\mathrm{B}T)}{n_\mathrm{\hi}\sigma_\mathrm{\hi}(24.59\mathrm{eV}+k_\mathrm{B}T)+n_\mathrm{\hei}\sigma_\mathrm{\hei}(24.59\mathrm{eV}+k_\mathrm{B}T)}\,,
\end{equation}
is the fraction of photons with energy higher than 24.59 eV emitted during Helium recombination that ionize hydrogen. The quantity $p$ is the fraction of ionizing photons generated by helium recombinations that are absorbed on the spot \citep{2006agna.book.....O}. 
Finally
$\tau_\nu$ is the optical depth:
\begin{equation}\label{eq:tau}
    \dfrac{d\tau_\nu}{dr}= 
\begin{cases}
    n_\mathrm{\hi}\sigma_\mathrm{\hi}(\nu),& \text{if } 13.6\leq h\nu/[\mathrm{eV}]\leq 24.59\\
    n_\mathrm{\hi}\sigma_\mathrm{\hi}(\nu)+n_\mathrm{\hei}\sigma_\mathrm{\hei}(\nu),              & \text{if } h\nu/[\mathrm{eV}]> 24.59.
\end{cases}
\end{equation}\par 
Assuming hydrogen is fully ionized throughout the \hii\ region, the  volume of the \hii\ region, $V_\mathrm{\hii}$, and radius, $R_\mathrm{\hii}$, can be roughly estimated as:
\begin{equation}\label{eq:VHII_estimate}
    Q_\mathrm{\hi}=\tilde{V}_\mathrm{\hii}\alpha_\mathrm{B,\hii}(T)n_\mathrm{H}^2=\dfrac{4\pi}{3}\tilde{R}_\mathrm{\hii}^3\alpha_\mathrm{B,\hii}(T)n_\mathrm{H}^2\,.
\end{equation}
Here we mark the \hii\ region volume and length scale with a tilde because Eq.~(\ref{eq:VHII_estimate}) is, again, an estimate. Specifically, Eq.~(\ref{eq:VHII_estimate}) assumes that hydrogen is completely ionized at $r\leq \tilde{R}_\mathrm{\hii}$. This is not a good assumption for very soft incident spectra, in which case the ISM transitions more smoothly from a fully ionized to a fully neutral phase.\par 
Throughout, Eq.~(\ref{eq:VHII_estimate}) and our approach assume that
all of the ionizing photons produced by each stellar particle are absorbed within a local \hii\ region surrounding the stellar particle in question. Otherwise, Eq.~(\ref{eq:VHII_estimate}) should be adjusted
with $(1-f_\mathrm{esc,loc})Q_\mathrm{\hi}$ on the left-hand side to account
for the fraction of ionizing photons that escape from the local \hii\ region, denoted here by $f_\mathrm{esc,loc}$. Some of the ionizing photons should, in fact, escape the galaxy entirely and ionize atoms in the intergalactic medium (IGM), while ionizing photons may also be consumed by neutral gas further away in the galaxy. We do not model these non-local effects in our calculations and treat each \hii\ region independently. Although this assumption is imperfect, note that the average
escape fraction into the IGM is likely small at these redshifts, perhaps $f_{\rm esc} \sim 0.1-0.2$, so on average $\sim 80-90\%$
of the ionizing photons are absorbed within a galaxy \citep[e.g.][]{2012ApJ...751...70V,2016MNRAS.461.3683I,2016Natur.529..178I,2017A&A...602A..18G,2018ApJ...869..123S}. The main simplification of our approach is to assume that the absorbed photons are consumed locally, and so it may mis-estimate the precise spatial distribution of the ionized gas. A small refinement might be to adopt a global average escape fraction and apply it to the \hii\ region around each stellar particle, but this would be a relatively small correction given the small escape fractions suggested by current observations. More challenging is to account for the non-local effects, which we leave to possible future work.\par
We then assume boundary conditions $n_\mathrm{\hii}=n_\mathrm{H}$, $n_\mathrm{\heii}=n_\mathrm{He}$ at radius $r=\tilde{R}_\mathrm{\hii}/100$, and solve Eqs.~(\ref{eq:HHebalance})--(\ref{eq:tau}) numerically to derive the radial profiles of $n_\mathrm{\hi}$ and $n_\mathrm{\hei}$ throughout the \hii\ region. We define the \hii\ region boundary $R_\mathrm{\hii}$ as the radius where $n_\mathrm{\hi}$ first surpasses $0.5n_\mathrm{H}$. We note that the presence of helium has no significant influence on the $n_\mathrm{\hi}$ radial profile. Therefore, to speed up the calculation we first ignore helium when solving the hydrogen ionization--recombination balance equation. With the derived $n_\mathrm{\hi}$ in each radial bin, we then solve for the helium ionization--recombination balance equation. The radial bin width is selected adaptively so that $n_\mathrm{\hi}$ and $n_\mathrm{\hei}$ vary by no more than 10\% between two adjacent bins. The \hii\ region volume is given by:
\begin{equation}
    V_\mathrm{\hii}=4\pi\int_\mathrm{\tilde{R}_\mathrm{\hii}/100}^{R_\mathrm{\hii}}r^2\dfrac{n_\mathrm{\hii}}{n_\mathrm{H}}dr\,.
\end{equation}\par
We then move on to derive the radial profiles of the \oiii\ and \oii\ number densities through solving the following ionization--recombination balance equations (including charge exchange reactions)\footnote{The time-scales for \oiii\ and \oii\ case B recombiantions are about 90 years and 800 years, much shorter than the O-star lifetimes and so assuming ioniziation equilibrium should be
an excellent assumption.}:
\begin{equation}\label{eq:OIIOIII_balance}
\begin{split}
    \dfrac{n_\mathrm{\oii}}{4\pi r^2}&\int_{\nu_\mathrm{\oii}}^\infty\dfrac{L_\nu}{h\nu}\sigma_\mathrm{\oii}(\nu)e^{-\tau_\nu}d\nu=\\
    &n_\mathrm{\oiii}n_e\alpha_\mathrm{B,\oiii}+n_\mathrm{\oiii}n_\mathrm{\hi}\delta_\mathrm{\oii}'\,,
\end{split}
\end{equation}
\begin{equation}\label{eq:OIOII_balance}
\begin{split}
    \dfrac{n_\mathrm{\oi}}{4\pi r^2}&\int_{\nu_\mathrm{\oi}}^\infty\dfrac{L_\nu}{h\nu}\sigma_\mathrm{\oi}(\nu)e^{-\tau_\nu}d\nu+n_\mathrm{\hii}n_\mathrm{\oi}k_{0r}=\\
    &n_\mathrm{\oii}n_e\alpha_\mathrm{B,\oii}+n_\mathrm{\hi}n_\mathrm{\oii}(k_0+k_1+k_2)\,,
\end{split}
\end{equation}
\begin{equation}\label{eq:O_tot}    n_\mathrm{\oiii}+n_\mathrm{\oii}+n_\mathrm{\oi}=n_\mathrm{O}\,.
\end{equation}      
Eqs.~(\ref{eq:OIIOIII_balance}) and (\ref{eq:OIOII_balance}) describe the ionization--recombination balance between \oiii\ and \oii, and \oii\ and \oi, respectively. Eq.~(\ref{eq:O_tot}) assumes that oxygen will only take the form of \oiii, \oii, or \oi\ in the \hii\ region, with the number density of oxygen characterized by the gas-phase metallicity $n_\mathrm{O}=10^{-3.31}n_\mathrm{H}(Z/Z_\odot)$. In the case of hard incident spectra, the UV photons can ionize oxygen into higher ionization states. However, in the FSPS models we consider here, highly ionized oxygen ions occupy a very limited volume near the centre of the \hii\ region, and have negligible influence on the overall volumes of the \oiii\ and \oii\ regions. The \oiii\ and \oii\ region volumes are:
\begin{equation}\label{eq:VOIIIVOII}
    \begin{split}
        V_X=4\pi\int_{\Tilde{R}_\mathrm{\hii}/100}^{R_\mathrm{\hii}}r^2\dfrac{n_X}{n_\mathrm{O}}dr\,.
    \end{split}
\end{equation}\par
Here $X$ specifies \oiii\ or \oii.\par
\begin{figure*}
    \centering
    \includegraphics[width=1\textwidth]{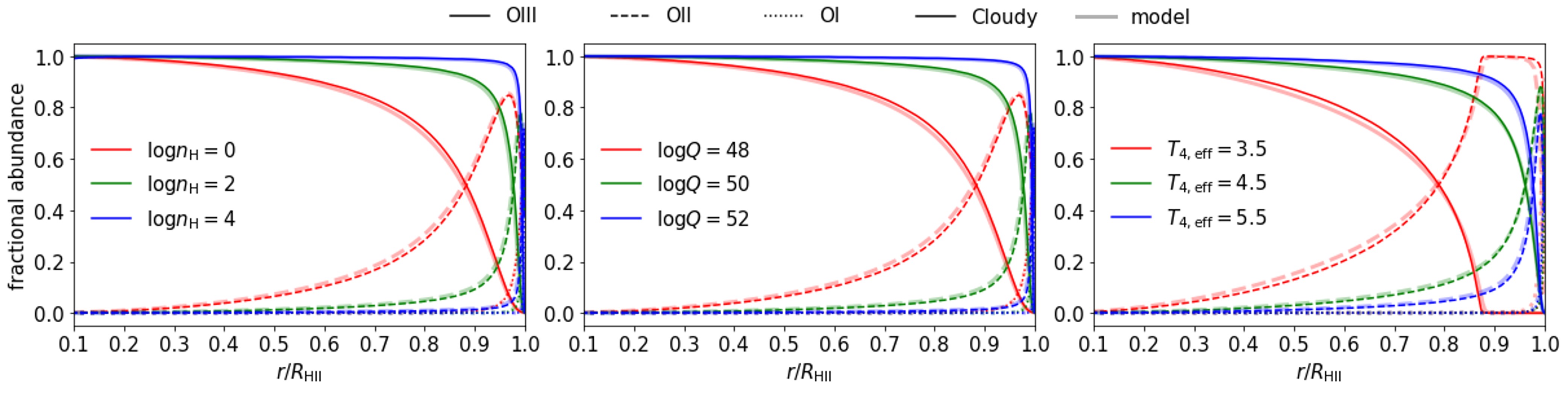}
    \caption{Fractional \oiii\ (solid), \oii\ (dashed), and \oi\ (dotted)  abundances as calculated by \textsc{Cloudy} (thin solid) and our model (thick transparent) under variations in the gas density $n_\mathrm{H}/[\mathrm{cm}^{-3}]$ (left), hydrogen ionizing photon generation rate $Q_\mathrm{\hi}/[\mathrm{s}^{-1}]$ (middle), and the effective temperature of a blackbody characterizing the spectral hardness (right). We have fixed the gas temperature of the \hii region to $T=10^4$,K. In the left panel, we fix the blackbody radiation spectrum strength $\log Q_\mathrm{\hi}/[\mathrm{s}^{-1}]=50$ and shape $T_\mathrm{4,eff}=T_\mathrm{eff}/10^4\,\mathrm{K}=5.5$, and only vary $n_\mathrm{H}$. For the middle panel, we fix $\log n_\mathrm{H}/[\mathrm{cm}^{-3}]=2$, $T_\mathrm{4,eff}=5.5$, and vary $Q_\mathrm{\hi}$. In the right panel $\log n_\mathrm{H}/[\mathrm{cm}^{-3}]=2$ and $\log Q_\mathrm{\hi}/[\mathrm{s}^{-1}]=50$ are fixed, but the spectral shape varies. Our model agrees with \textsc{Cloudy} in all cases. }\label{fig:OIIIProfile}
\end{figure*}\par

\begin{figure}
    \centering
    \includegraphics[width=0.49\textwidth]{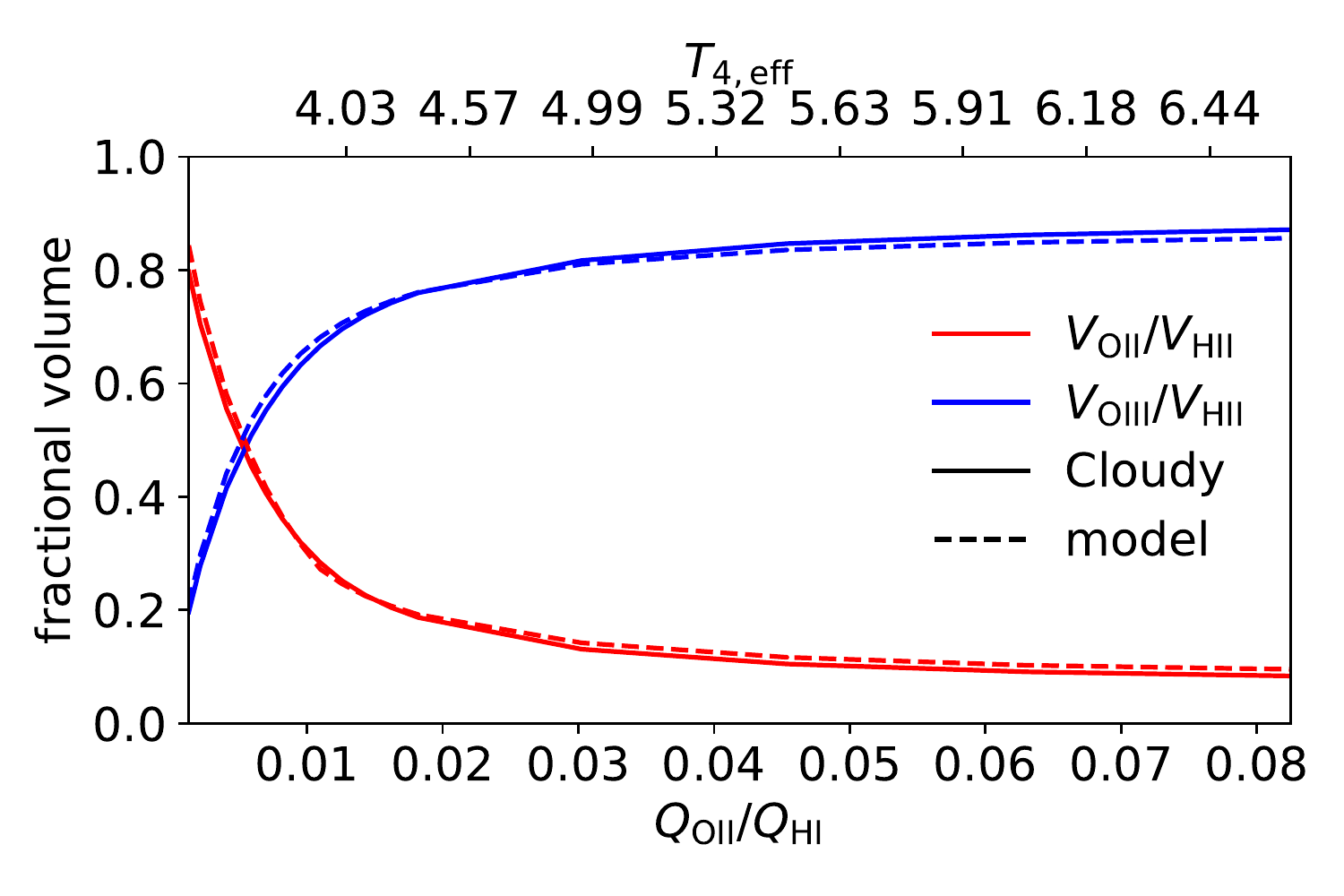}
    \caption{\oii\ (red) and \oiii\ region (blue) fractional volumes as calculated by \textsc{Cloudy} (solid) and our model (dashed) under variations in the shape of the incident radiation spectrum. We have fixed the gas temperature to $T=10^4$\,K, the gas density at $n_\mathrm{H}=100$\,cm$^{-3}$, and the incident spectrum strength to $Q_\mathrm{\hi}=10^{50}$\,s$^{-1}$. The model prediction is in good agreement with \textsc{Cloudy}.
    }\label{fig:ratioV_T}
\end{figure}

We compare this simple \hii, \oiii, and \oii\ region volume model with \textsc{Cloudy} calculations and show the results in Figures~\ref{fig:OIIIProfile}~and~\ref{fig:ratioV_T}. In Figure~\ref{fig:OIIIProfile} we test how the \oiii\ and \oii\ radial distributions are influenced by the gas density in the \hii\ region, and with variations in the strength and shape of the incident spectrum. We fix the gas temperature of the \hii\ region at $T=10^4$,K. In the left and middle panels of Figure~\ref{fig:OIIIProfile} we fix the incident spectral shape to a blackbody with an effective temperature of $T_\mathrm{eff}=55\,000$\,K, which is close to the \textsc{starburst99} spectrum for a continuous SFR model with an age of 6\,Myr \citep{2020MNRAS.499.3417Y}. In the left panel, we set the hydrogen ionization rate to $Q_\mathrm{\hi}=10^{50}$\,s$^{-1}$ and only vary the gas density, while in the middle panel we fix the gas density to $n_\mathrm{H}=100$\,cm$^{-3}$ and only vary $Q_\mathrm{\hi}$. In the right panel, we set $n_\mathrm{H}=100$\,cm$^{-3}$, $Q_\mathrm{\hi}=10^{50}$\,s$^{-1}$, and vary the effective temperature of the blackbody radiation spectrum, $T_\mathrm{4,eff}=T_\mathrm{eff}/10^4$\,K. The \oiii, \oii, and \oi\ radial distributions throughout the \hii\ region are presented as the solid, dashed, and dotted lines, respectively. The \textsc{Cloudy} and model predictions are given by thin solid and thick transparent curves, respectively. In every case, our simple model agrees well with \textsc{Cloudy}.\par 

The ratio between the volumes of the \oiii\ and \hii\ regions, $V_\mathrm{\oiii}/V_\mathrm{\hii}$, increases with $n_\mathrm{H}$ and $Q_\mathrm{\hi}$. This can be shown through combining Eqs.~(\ref{eq:tau})~and~(\ref{eq:VHII_estimate}):
\begin{equation}
\begin{split}
    \dfrac{d\tau_\nu}{d(r/R_\mathrm{\hii})}&=n_\mathrm{\hi}\sigma_\mathrm{\hi}(\nu)R_\mathrm{\hii}\approx n_\mathrm{\hi}\sigma_\mathrm{\hi}(\nu)\tilde{R}_\mathrm{\hii}\\
    &\propto x_\mathrm{\hi}n_\mathrm{H}^{1/3}Q_\mathrm{\hi}^{1/3}\,,
\end{split}
\end{equation}
here $x_\mathrm{\hi}=n_\mathrm{\hi}/n_\mathrm{H}$ is the fractional abundance of neutral hydrogen. At a fixed optical depth radial derivative, a greater $n_\mathrm{H}$ or $Q_\mathrm{\hi}$ will lead to smaller \hi, \oii, and \oi\ fractional abundances and to a sharper \oiii\ region edge. In the right panel of Figure~\ref{fig:OIIIProfile} the \oiii\ region fails to extend to the \hii\ region edge for $T_\mathrm{4,eff}=3.5$. This is because this very soft spectrum creates an \heii\ region that is smaller than the \hii\ region. Since the ionization energy for generating \oiii\ is 35.12 eV, higher than the ionization energy 24.59 eV for \hei\ $\rightarrow$ \heii, the \oiii\ region cannot extend beyond the \heii\ region.\par 

In Figure~\ref{fig:ratioV_T} we fix the \hii\ region gas density at $n_\mathrm{H}=100$\,cm$^{-3}$, the temperature to $T=10^4$\,K, and the incident spectrum amplitude to $Q_\mathrm{\hi}=10^{50}$\,s$^{-1}$, while varying the shape of the incident spectrum characterized by $T_\mathrm{4,eff}$. We compare the ratio between the \oii\ and \hii\ region volumes (red) and the \oiii\ versus \hii\ region volumes (blue), as predicted by \textsc{Cloudy} (solid) and our model (dashed). Our model is again in excellent agreement with \textsc{Cloudy}. \par

\par

\subsection{5-level population abundance}
In this subsection we introduce models for calculating the H$\alpha$, H$\beta$, [\oiii], and [\oii] line luminosities. With the simple picture introduced at the beginning of Section~\ref{subsec:volume} it is straightforward to determine the H$\alpha$ and H$\beta$ line luminosities \citep{2011piim.book.....D}:
\begin{equation}\label{eq:LHalphabeta}
\begin{split}
    L_Y =h\nu_Y\dfrac{\alpha_{\mathrm{B},Y}(T)}{\alpha_\mathrm{B,\hii}(T)}Q_\mathrm{\hi}\,.
\end{split}
\end{equation}
Here $Y$ specifies H$\alpha$ or H$\beta$ line emission.  This relation follows from ionization equilibrium, with photoionizations balancing recombinations, provided that each ionizing photon is absorbed locally within the HII region (as mentioned previously). The factor $\alpha_{\mathrm{B},Y}/\alpha_\mathrm{B,\hi}$ is the fraction of case-B hydrogen recombinations that lead to the emission of the H$\alpha$ or H$\beta$ line emission.\par

\begin{figure}
    \centering
    \includegraphics[width=0.49\textwidth]{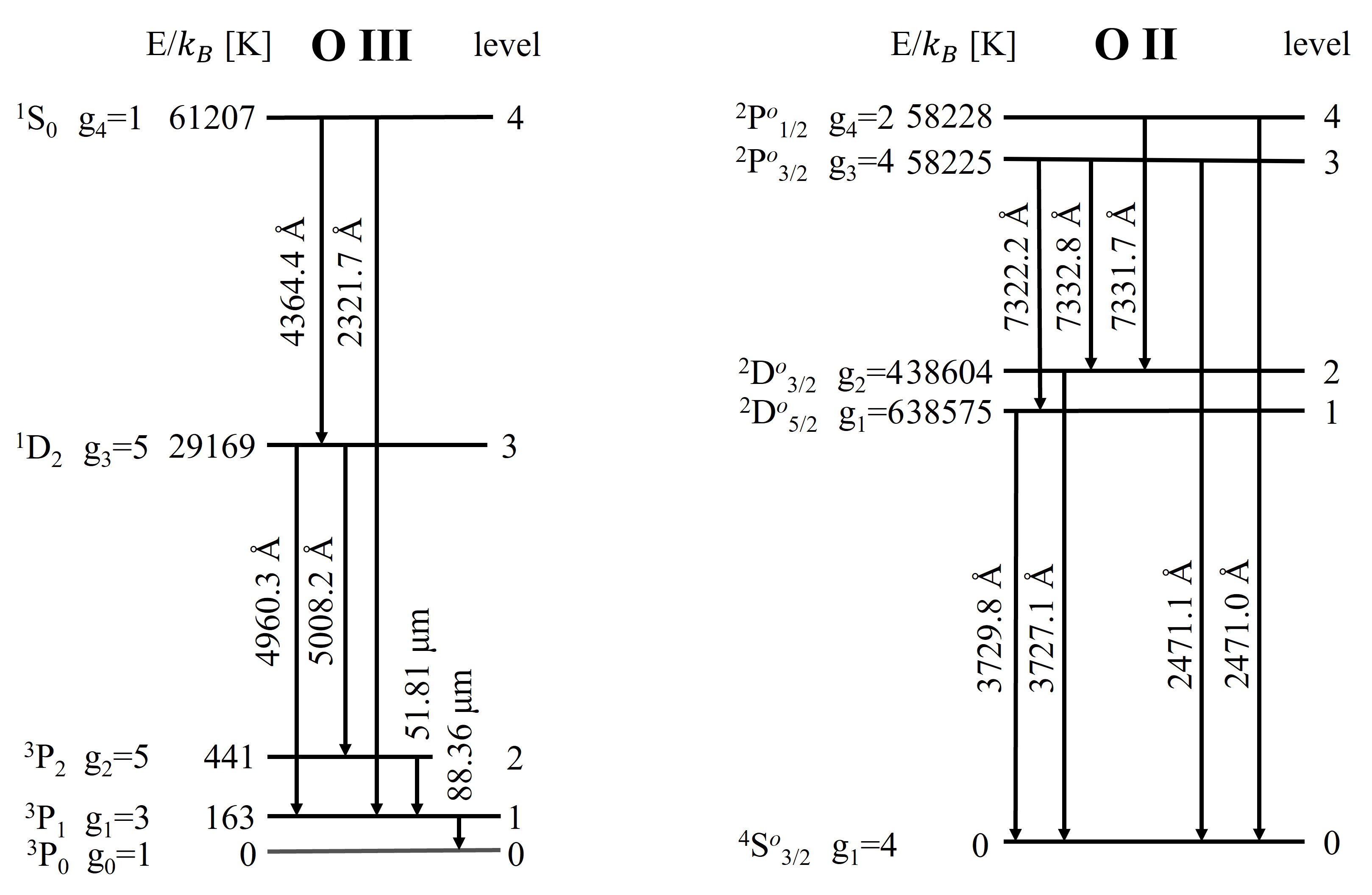}
    \caption{The lowest five energy levels of the \oiii\ and \oii\ ions. The columns, from left to right, show the spectroscopic term, degeneracy, and energy relative to the ground state for the corresponding energy level. The wavelengths of the radiative transitions covered in this work are also specified. Adapted from \protect\cite{2011piim.book.....D}.}\label{fig:ion_structure}
\end{figure}

In Figure~\ref{fig:ion_structure}, we show the energy levels and transitions for the ground electron configurations of \oiii\ ($1s^22s^22p^2$) and \oii\ ($1s^22s^22p^3$) ions. The luminosity in any [\oiii] or [\oii] line emitted in a transition between energy levels  $i\rightarrow j$ can be modeled as:
\begin{equation}\label{eq:LOIIIOII}
\begin{split}
    L^X_{ij}&=h\nu^X_{ij}A^X_{ij}\int_0^{R_\mathrm{\hii}}n^X_i4\pi r^2dr\\
    &=h\nu^X_{ij}A^X_{ij}\dfrac{n^X_i}{n_{X}}n_\mathrm{O}\int_0^{R_\mathrm{\hii}}\dfrac{n_{X}}{n_\mathrm{O}}4\pi r^2dr\\
    &=h\nu^X_{ij}A^X_{ij}\dfrac{n^X_{i}}{n_{X}}n_\mathrm{O}V_{X}\\
    &\approx h\nu^X_{ij}A^X_{ij}\dfrac{n^X_i}{n_{X}}n_\mathrm{O}\dfrac{Q_\mathrm{\hi}}{\alpha_\mathrm{B,\hii}(T)n_\mathrm{H}^2}\dfrac{V_{X}}{\tilde{V}_\mathrm{\hii}} \,.
\end{split}
\end{equation}
Here $X$ specifies \oiii\ or \oii. The volume correction factors $V_\mathrm{\oiii}/\tilde{V}_\mathrm{\hii}$ and $V_\mathrm{\oii}/\tilde{V}_\mathrm{\hii}$ are given by Eq.~(\ref{eq:VOIIIVOII}). To maintain high computational efficiency, we consider only the radial variations in free electron density when solving for volume correction factors, but will assume $n_\mathrm{e}=n_\mathrm{H}+n_\mathrm{He}$ when modelling the \oiii\ and \oii\ ion level populations. In principle, neglecting the full radial dependence of $n_e$ is imperfect. In practice, however, the line luminosities depend directly on $n_e$ only at high densities relative to the critical densities of the lines of interest. The high density \hii\ regions are almost entirely sourced by young stellar populations, which have hard ionizing spectra. In this case, the ionization fraction has little radial dependence and so our approximation should be an excellent one.  \hii\ regions with softer radiation fields do show more radial dependence in their ionization fractions, but these regions tend to be at lower density and so the line luminosities become independent of $n_e$ in this limit. This further justifies our approach.
\par
\par
Assuming that the \oiii\ and \oii\ ions have achieved a steady-state where the level population abundances do not vary with time:
\begin{equation}\label{eq:levelPop}
    \begin{split}
        \dfrac{dn_1}{dt}=R_{01}n_0&+R_{21}n_2+R_{31}n_3+R_{41}n_4\\
        &-(R_{10}+R_{12}+R_{13}+R_{14})n_1=0\,,\\
        \dfrac{dn_2}{dt}=R_{02}n_0&+R_{12}n_1+R_{32}n_3+R_{42}n_4\\
        &-(R_{20}+R_{21}+R_{23}+R_{24})n_2=0\,,\\
        \dfrac{dn_3}{dt}=R_{03}n_0&+R_{13}n_1+R_{23}n_2+R_{43}n_4\\
        &-(R_{30}+R_{31}+R_{32}+R_{34})n_3=0\,,\\
        \dfrac{dn_4}{dt}=R_{04}n_0&+R_{14}n_1+R_{24}n_2+R_{34}n_3\\
        &-(R_{40}+R_{41}+R_{42}+R_{43})n_4=0\,.\\
    \end{split}
\end{equation}
Here we have omitted the \oiii\ and \oii\ superscripts for simplicity. The quantity $R_{ij}$ is the rate at which \oiii\ or \oii\ ions transition from level $i$ to level $j$ through spontaneous decay or collisional excitation/de-excitation:
\begin{equation}
R_{ij}=
    \begin{cases}
    n_ek_{ij}(T)+A_{ij},& \text{if } i>j\\
    n_ek_{ij}(T),       & \text{if } i<j.
\end{cases}
\end{equation}
We can then solve for the relative level populations, $\{n_1/n_0\,,n_2/n_0\,,n_3/n_0\,,n_4/n_0\}$. Given $\Sigma_{i=0}^4n^X_{i}=n_{X}$ we can derive $n^X_{i}/n_{X}$ for all five energy levels.\par
\begin{figure*}
    \centering
    \includegraphics[width=1\textwidth]{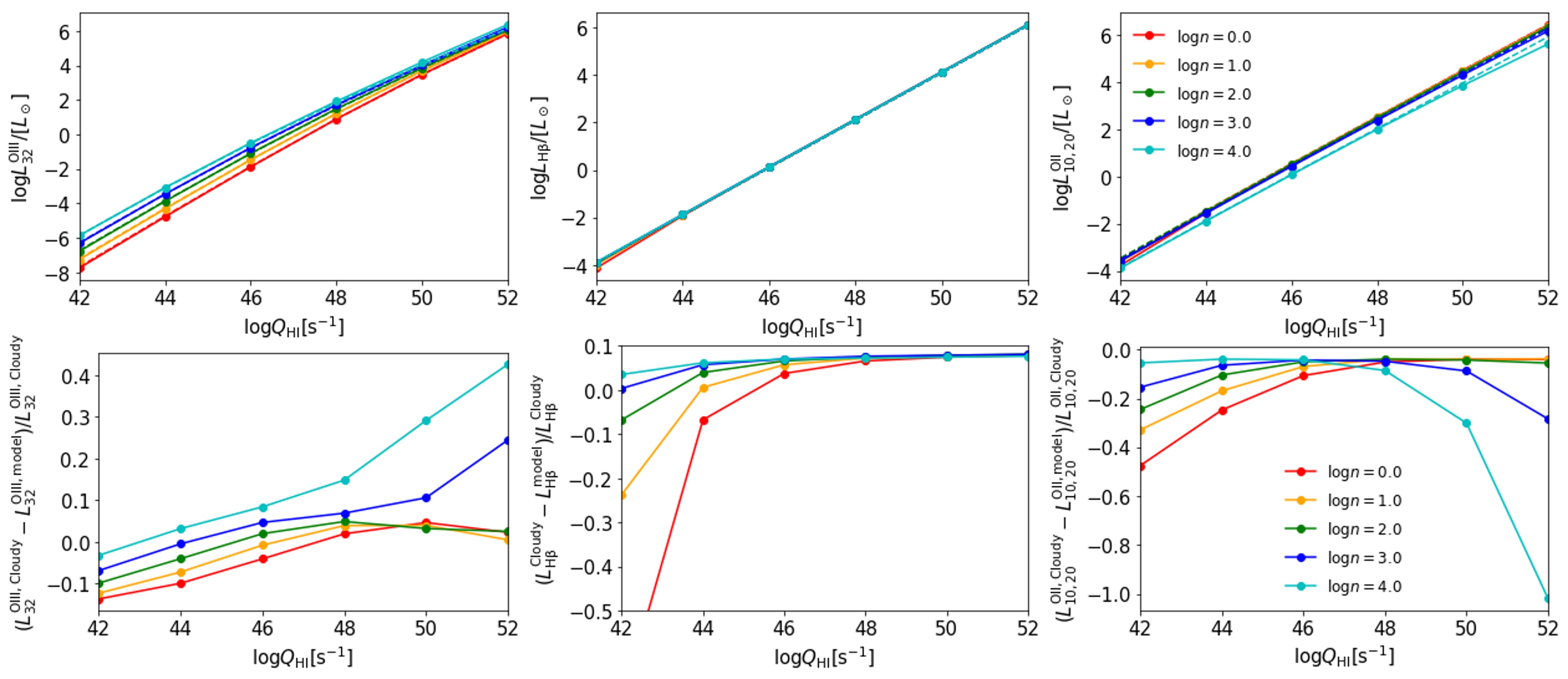}
    \caption{Comparisons between our model Eqs.~(\ref{eq:LHalphabeta})~and~(\ref{eq:LOIIIOII}) and \textsc{Cloudy} simulation results for [\oiii] 5008$\angstrom$(left column), H$\beta$ (middle column), and [\oii] 3727,30$\angstrom$(right column) lines. In the top row, the model predictions are presented as dashed curves while \textsc{Cloudy} solutions are shown as solid lines. We show the fractional difference between our model and \textsc{Cloudy} in the bottom row. For each test we have fixed the gas temperature to $T=10^4$\,K and the incident spectrum shape to $T_\mathrm{4,eff}=3$. Our model is generally in good agreement with \textsc{Cloudy}. Our \oiii\ and \oii\ models become inaccurate for dense gas environments facing soft incident spectra with high $Q_\mathrm{\hi}$. Our model also overestimates $L_\mathrm{H\beta}$ at low $Q_\mathrm{\hi}$. We will show in Section~\ref{sec:FIRE} that the parameter space where our model becomes inaccurate is largely irrelevant for galaxy zoom-in simulations.}\label{fig:softSpec}
\end{figure*}
\begin{figure*}
    \centering
    \includegraphics[width=1\textwidth]{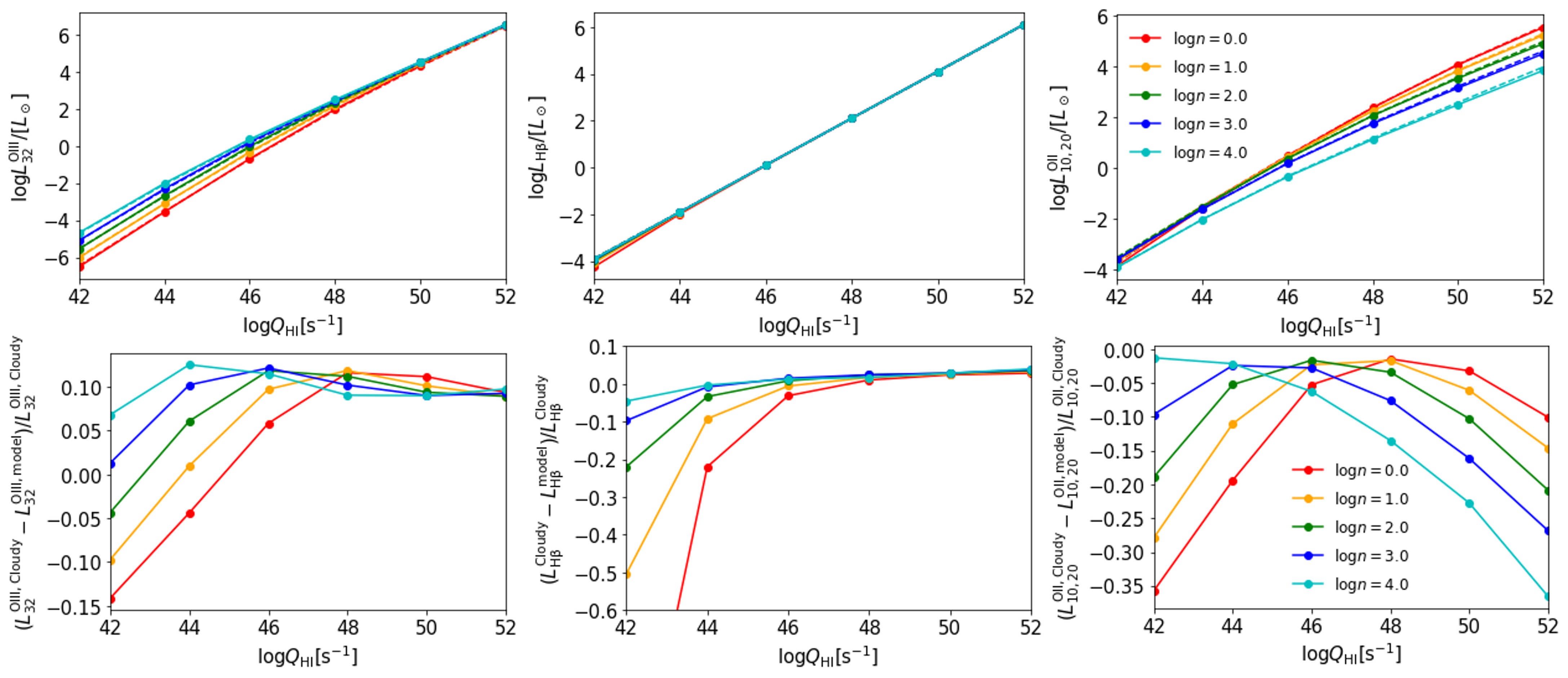}
    \caption{Similar to Figure~\ref{fig:softSpec}, but here the incident spectral shape is fixed as $T_\mathrm{4,eff}=5.5$. The agreement between our model and \textsc{Cloudy} becomes better under a harder spectrum. However, our model still underestimates $L_\mathrm{H\beta}$ at low $Q_\mathrm{\hi}$. We will show in the Section~\ref{sec:FIRE} that stellar particles with $Q_\mathrm{\hi}\lesssim 10^{49}$\,s$^{-1}$ make a negligible contribution to the overall H$\beta$ line luminosity.}\label{fig:hardSpec}
\end{figure*}

Now we have all the ingredients needed to calculate the line luminosities using Eqs.~(\ref{eq:LHalphabeta})~and~(\ref{eq:LOIIIOII}). In Figures~\ref{fig:softSpec}~and~\ref{fig:hardSpec} we compare our model with \textsc{Cloudy} simulation results for [\oiii] 5008$\angstrom$(left column), $H\beta$ (middle column), and [\oii] 3727,30$\angstrom$(right column) lines under a relatively soft ($T_\mathrm{4,eff}=3$) and hard ($T_\mathrm{4,eff}=5.5$) spectrum, respectively. In each case, we fix the gas-phase metallicity to $Z=0.2\,Z_\odot$ and gas temperature to $T=10^4$\,K. In the first row of each figure we compare our model predictions (dashed) with \textsc{Cloudy} (solid) under a variety of gas densities. In the second row of each figure we show the fractional difference between our model and \textsc{Cloudy}. The agreement between our model and \textsc{Cloudy} is generally better than 30 per cent in the range $42\leq \log Q_\mathrm{\hi}/[\mathrm{s}^{-1}]\leq52$ and $0\leq \log n_\mathrm{H}/[\mathrm{cm}^{-3}]\leq4$. We will show in Section~\ref{sec:FIRE} that these ranges are most relevant for FIRE stellar particles and their surrounding ISM properties. The inaccuracy in our model is dominated by the assumption that $n_\mathrm{e}=n_\mathrm{H}$, which is inaccurate for dense gas environments facing a soft incident spectrum. We will show in Section~\ref{sec:FIRE} that the FIRE star particles with $Q_\mathrm{\hi}\gtrsim10^{50}\,\hbox{s}^{-1}$ are all very young stars with hard radiation spectra and are embedded in dense ionized gas. Our model also significantly underestimates $L_\mathrm{H\beta}$ for $Q_\mathrm{\hi}\lesssim 10^{44}$\,s$^{-1}$, where Eq.~(\ref{eq:VHII_estimate}) becomes invalid. We will show later that the H$\beta$ line is mainly contributed by young stars with $Q_\mathrm{\hi}\gtrsim 10^{49}$\,s$^{-1}$. In conclusion, the parameter space where our model becomes inaccurate is largely irrelevant for current galaxy zoom-in simulations.\par

\par

\par

\subsection{Line luminosity in the low gas density limit}\label{subsec:LatlownH}
The critical density of a line is defined as $n_{\mathrm{crit},i}=\Sigma_{j<i}A_{ij}/\Sigma_{j<i}k_{ij}$. In the low gas density limit where $n_e\approx n_\mathrm{H}\ll n_{\mathrm{crit},i}$, solutions of the level population balance equations, i.e. Eq.~(\ref{eq:levelPop}), can be greatly simplified as then the collisional de-excitation terms $k_{ij}n_e$ can be ignored in the denominator. Detailed \oiii\ ion level populations under the low gas density limit have been derived in \cite{2020MNRAS.499.3417Y}. Below we summarize the critical densities and luminosity models for multiple lines observed by ALMA and \textit{JWST} in Eq.~(\ref{eq:model_lownH}).
Notice that except for the volume correction factors $V_\mathrm{\oiii}/\tilde{V}_\mathrm{\hii}$ and $V_\mathrm{\oii}/\tilde{V}_\mathrm{\hii}$, the remainder of the line emission model in the low gas density limit is independent of $n_\mathrm{H}$.

\clearpage

\begin{strip}
\begin{equation}\label{eq:model_lownH}
\begin{split}
    L_{10}^\mathrm{\oiii}&=\left(\dfrac{n_\mathrm{O}}{n_\mathrm{H}}\right)_\odot\dfrac{Z}{Z_\odot}(k_{01}^\mathrm{\oiii}+k_{02}^\mathrm{\oiii})h\nu_{10}^\mathrm{\oiii}\dfrac{Q_\mathrm{\hi}}{\alpha_\mathrm{B,\hii}}\dfrac{V_\mathrm{\oiii}}{\tilde{V}_\mathrm{\hii}}\,,\ \mathrm{for}\ n_\mathrm{H}\ll n_\mathrm{crit,1}^\mathrm{\oiii}=1.7\times10^3\ \mathrm{cm}^{-3},\\
    L_{21}^\mathrm{\oiii}&=\left(\dfrac{n_\mathrm{O}}{n_\mathrm{H}}\right)_\odot\dfrac{Z}{Z_\odot}k_{02}^\mathrm{\oiii}h\nu_{21}^\mathrm{\oiii}\dfrac{Q_\mathrm{\hi}}{\alpha_\mathrm{B,\hii}}\dfrac{V_\mathrm{\oiii}}{\tilde{V}_\mathrm{\hii}}\,,\ \mathrm{for}\ n_\mathrm{H}\ll n_\mathrm{crit,2}^\mathrm{\oiii}=3.8\times10^3\ \mathrm{cm}^{-3},\\
    L_{31}^\mathrm{\oiii}&=\dfrac{A_{31}^\mathrm{\oiii}}{A_{31}^\mathrm{\oiii}+A_{32}^\mathrm{\oiii}}\left(\dfrac{n_\mathrm{O}}{n_\mathrm{H}}\right)_\odot\dfrac{Z}{Z_\odot}k_{03}^\mathrm{\oiii}h\nu_{31}^\mathrm{\oiii}\dfrac{Q_\mathrm{\hi}}{\alpha_\mathrm{B,\hii}}\dfrac{V_\mathrm{\oiii}}{\tilde{V}_\mathrm{\hii}}\,,\ \mathrm{for}\ n_\mathrm{H}\ll n_\mathrm{crit,3}^\mathrm{\oiii}=7.4\times10^5\ \mathrm{cm}^{-3},\\
    L_{32}^\mathrm{\oiii}&=\dfrac{A_{32}^\mathrm{\oiii}}{A_{31}^\mathrm{\oiii}+A_{32}^\mathrm{\oiii}}\left(\dfrac{n_\mathrm{O}}{n_\mathrm{H}}\right)_\odot\dfrac{Z}{Z_\odot}k_{03}^\mathrm{\oiii}h\nu_{32}^\mathrm{\oiii}\dfrac{Q_\mathrm{\hi}}{\alpha_\mathrm{B,\hii}}\dfrac{V_\mathrm{\oiii}}{\tilde{V}_\mathrm{\hii}}\,,\ \mathrm{for}\ n_\mathrm{H}\ll n_\mathrm{crit,3}^\mathrm{\oiii}=7.4\times10^5\ \mathrm{cm}^{-3},\\
    L_{43}^\mathrm{\oiii}&=\dfrac{A_{43}^\mathrm{\oiii}}{A_{43}^\mathrm{\oiii}+A_{41}^\mathrm{\oiii}}\left(\dfrac{n_\mathrm{O}}{n_\mathrm{H}}\right)_\odot\dfrac{Z}{Z_\odot}k_{04}^\mathrm{\oiii}h\nu_{43}^\mathrm{\oiii}\dfrac{Q_\mathrm{\hi}}{\alpha_\mathrm{B,\hii}}\dfrac{V_\mathrm{\oiii}}{\tilde{V}_\mathrm{\hii}}\,,\ \mathrm{for}\ n_\mathrm{H}\ll n_\mathrm{crit,4}^\mathrm{\oiii}=2.6\times10^7\ \mathrm{cm}^{-3},\\
    L_{10}^\mathrm{\oii}&=\left(\dfrac{n_\mathrm{O}}{n_\mathrm{H}}\right)_\odot\dfrac{Z}{Z_\odot}k_{01}^\mathrm{\oii}h\nu_{10}^\mathrm{\oii}\dfrac{Q_\mathrm{\hi}}{\alpha_\mathrm{B,\hii}}\dfrac{V_\mathrm{\oii}}{\tilde{V}_\mathrm{\hii}}\,,\ \mathrm{for}\ n_\mathrm{H}\ll n_\mathrm{crit,1}^\mathrm{\oii}=3.6\times10^3\ \mathrm{cm}^{-3},\\
    L_{20}^\mathrm{\oii}&=\left(\dfrac{n_\mathrm{O}}{n_\mathrm{H}}\right)_\odot\dfrac{Z}{Z_\odot}k_{02}^\mathrm{\oii}h\nu_{20}^\mathrm{\oii}\dfrac{Q_\mathrm{\hi}}{\alpha_\mathrm{B,\hii}}\dfrac{V_\mathrm{\oii}}{\tilde{V}_\mathrm{\hii}}\,,\ \mathrm{for}\ n_\mathrm{H}\ll n_\mathrm{crit,2}^\mathrm{\oii}=3.8\times10^3\ \mathrm{cm}^{-3}.\\
\end{split}    
\end{equation}
\end{strip}

%%%%%%%%%%%%%%%%%%%%%%%%%%%%%%%%%%%%%%%%%%%%%%%%%%
\section{FIRE ISM emission post-processing}\label{sec:FIRE}
As a first application of our model, we present post-processed predictions for the ISM line emission from the 22 primary galaxies in the publicly available FIRE high-$z$ suite \citep{2018MNRAS.478.1694M,2022arXiv220206969W} at $z=6$. The initial baryonic particle masses for different zoom-in boxes range from $10^2$--$10^4\mathrm{M}_\odot$. Assuming a typical \hii\ region gas density of $n_\mathrm{H}\sim100$\,cm$^{-3}$, FIRE can resolve HII regions with length scales of $\sim3$--16\,pc, which is comparable to the size of the \hii\ region generated by O-stars residing in molecular clouds \citep[e.g.][]{2014MNRAS.442..694D}. Although the FIRE high-$z$ galaxy sample volume is small, the galaxy stellar mass range covers $10^6$--$10^{10}\mathrm{M}_\odot$. Therefore,
we can compare
post-processed predictions of the ISM line emission from
the FIRE high-$z$ suite with current \textit{JWST} measurements in the stellar mass range $10^7$--$10^{9}\mathrm{M}_\odot$.\par 
Within each FIRE primary galaxy we treat \hii\ regions sourced by individual stellar particles as line emitters. The gas properties for each \hii\ region are estimated from the gas particle neighbours around the corresponding stellar particle. Specifically, we use \mbox{\textsc{GizmoAnalysis}}\footnote{\url{http://ascl.net/2002.015}, first used in \cite{2016ApJ...827L..23W}.} to find the stellar mass centre of each FIRE high-$z$ primary galaxy. We then define $R_{90}$ as the radius that encloses 90 per cent of the stellar mass within 20\,kpc from the stellar mass centre, and treat all gas and stellar particles within $2R_{90}$ as baryonic components of the primary galaxy. Given the age, stellar mass, and stellar metallicity of each stellar particle in the primary galaxy, we model the stellar radiation through linear interpolation over a FSPS spectral lookup table\footnote{ \url{https://zenodo.org/record/6338462}} \citep{2010ascl.soft10043C}, assuming a Chabrier initial mass function \citep{2001ApJ...554.1274C}. To determine the gas environment 
in the \hii\ region
around each star particle, we find the nearest 32 gas particle neighbours to each stellar particle and define the \hii\ region density, $n_\mathrm{H}$, and metallicity, $Z$, as averages over all of the gas particle neighbours. Note that the choice of averaging over the nearest 32 gas particles is arbitrary. In particular, this procedure neglects the non-local ionization effects mentioned earlier. Nevertheless, our scheme still captures the fact that \hii\ regions will tend -- with all other things being equal -- to be smaller in denser environments and larger in more rarefied portions of a galaxy. Although in future work it may be interesting to develop more sophisticated non-local photoionization models here, this may be more appropriate for higher resolution simulations since the FIRE high-$z$ suite only marginally resolves individual \hii\ regions.\par
\begin{figure}
    \centering
    \includegraphics[width=0.49\textwidth]{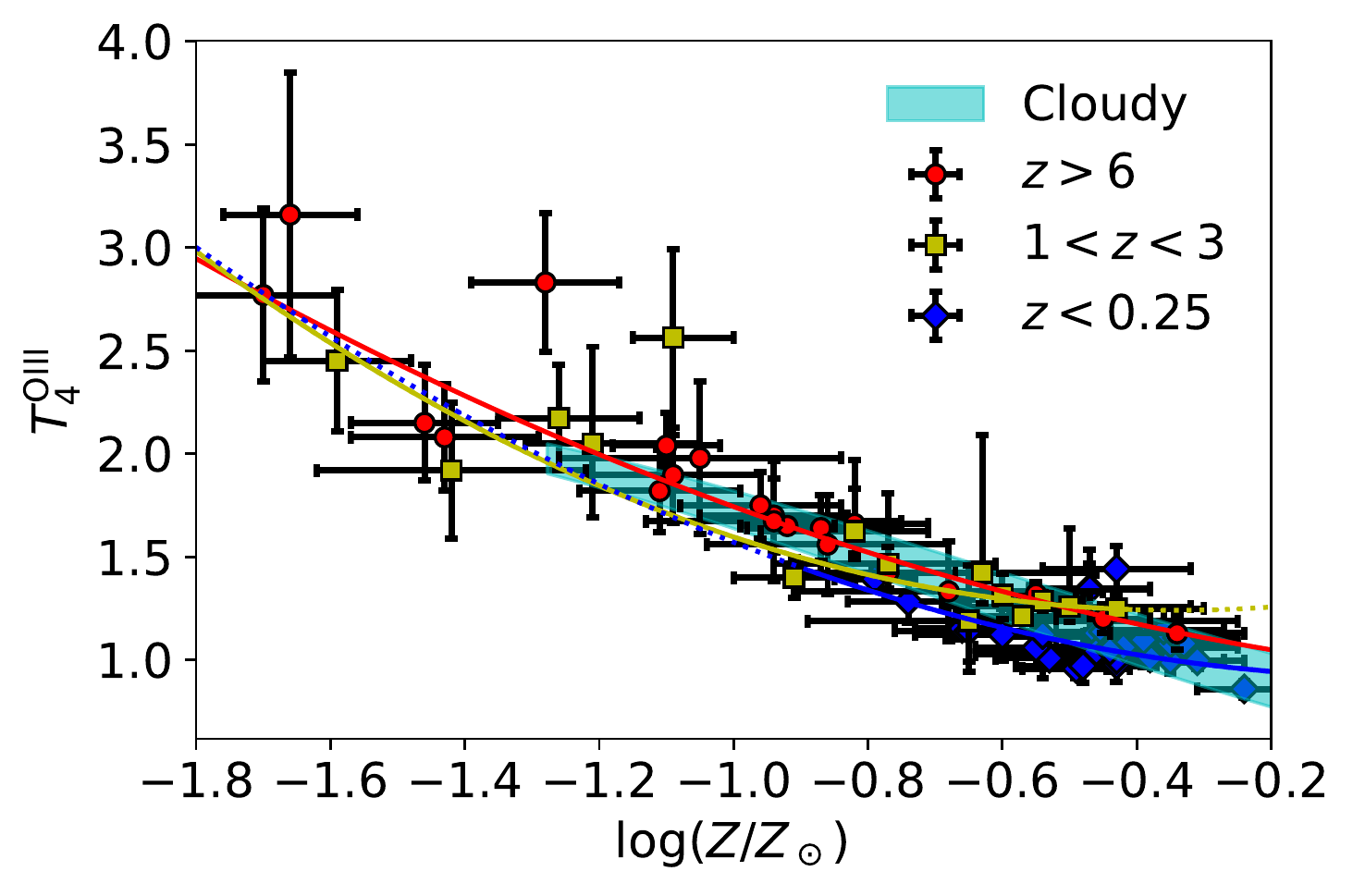}
    \caption{Redshift evolution of the \oiii\ region temperature versus metallicity relation. All data points are measured from the direct $T_e$ method. The red points are high-z measurements from \protect\cite{2023arXiv230603120L,2023arXiv230112825N,2023arXiv230308149S,2023MNRAS.518..425C}. The yellow symobols show measurements at the cosmic noon \protect\citep{2018MNRAS.481.3520P,2023arXiv230603120L,2023arXiv230308149S}. The blue points are local measurements from \protect\cite{2020A&A...634A.107Y}. Solid curves of the corresponding colours show the best-fit polynomial relations. The dotted curves extrapolate the TZRs outside of their callibrated metallicity range. 
    We find that the evolution in the TZR is generally less than 2000 K over wide redshift ranges. Assuming thermal equilibrium is achieved everywhere within the ISM, the TZR from \textsc{Cloudy} over $50\leq\log Q_\mathrm{\hi}/\mathrm{[s^-1]}\leq56$, $0\leq\log n_\mathrm{H}/\mathrm{[cm^{-3}]}\leq4$ agrees very well with the measurements at all redshifts (cyan band).}\label{fig:TZz}
\end{figure}

 
In this work we do not adopt the \hii\ region gas temperatures simulated by FIRE, but instead assume an observationally motivated temperature versus metallicity relation (TZR). The \oiii\ region gas temperatures at various redshifts have been constrained by determinations of their $L_{43}^\mathrm{\oiii}/L_{32}^\mathrm{\oiii}$ luminosity ratios (the ``direct $T_e$ method''). We collect direct $T_e$ measurements from the EoR \citep{2023arXiv230603120L,2023arXiv230112825N,2023arXiv230308149S,2023MNRAS.518..425C}, cosmic noon \citep{2018MNRAS.481.3520P,2023arXiv230603120L,2023arXiv230308149S}, and the local universe \citep{2020A&A...634A.107Y}. The \oiii\ region TZRs are shown in Figure~\ref{fig:TZz}. The red circles, yellow squares, and blue diamonds show data at $z>6$, $1<z<3$, and $z<0.25$, respectively. The best-fit TZRs at these three different redshift intervals are presented by curves of the corresponding colours:
\begin{equation}\label{eq:TZ}
    \begin{split}
        T_4^\mathrm{\oiii}&=0.40\log Z^2-0.39\log Z+0.96\,,\ z>6\,,\\
        T_4^\mathrm{\oiii}&=0.82\log Z^2+0.56\log Z+1.3\,,\ 1<z<3\,,\\
        T_4^\mathrm{\oiii}&=0.63\log Z^2-0.029\log Z+0.91\,,\ z<0.25\,.
    \end{split}
\end{equation}
Here the metallicity $Z$ is in units of solar metallicity and $T_4^\mathrm{\oiii}$ is the \oiii\ region gas temperature in units of $10^4$\,K. We find that over such a wide redshift range, the TZR variation is mostly less than 2000\,K at a fixed metallicity, which is much less than the typical \hii\ region gas temperature. Such a small temperature variation would cause tiny changes in the line ratios among [\oiii], [\oii], and hydrogen Balmer lines. Therefore we conclude that current direct $T_e$ measurements suggest a universal TZR. In this work we will use the best-fit TZR at $z>6$ to derive strong line diagnostics. Swiching to the best-fit TZR at other redshifts will not alter any of our main conclusions.\par 
It is expected that the temperature of the ionized ISM decreases with metallicity owing to the enhanced cooling of metal-enriched gas. Indeed, we find that the observed TZR trend can be nicely reproduced by determining the balance between heating and cooling processes in thermal equilibrium. \cite{2020MNRAS.499.3417Y} developed an \hii\ region temperature model fit to \textsc{Cloudy} \citep{2017RMxAA..53..385F} simulation results, where thermal equilibrium is assumed to be established everywhere within the 1D ISM, assuming stellar radiation spectra from \textsc{starburst99} (see their Eq.~2). The model is calibrated over a wide ISM parameter space $50\leq\log Q_\mathrm{\hi}/\mathrm{[s^-1]}\leq56$, $0\leq\log n_\mathrm{H}/\mathrm{[cm^{-3}]}\leq4$, and $0.05\leq Z/[Z_\odot]\leq1$. Here $Q_\mathrm{\hi}$ is the hydrogen ionizing photon generation rate summed over all emitters within a galaxy, while $n_\mathrm{H}$ is the \hii\ region gas density. We show the TZR range given by this model within the calibrated parameter space as a cyan band in Figure~\ref{fig:TZz}. The \textsc{Cloudy} simulation results are fully consistent with the observational measurements. Whether this agreement provides evidence that thermal equilibrium is largely established within the ISM or is just a coincidence requires further studies.\par
We ignore temperature fluctuations within each galaxy and assume that the \hii\ and \oiii\ region gas temperatures are equal, i.e. $T_4^\mathrm{\hii}=T_4^\mathrm{\oiii}$. Finally, we adopt
empirical calibrations based on direct $T_e$ measurements towards 130 local galaxies to determine
the \oii\ temperatures from 
the \oiii\ ones. Specifically, we assume  \citep{2020A&A...634A.107Y}:
\begin{equation}
        T^\mathrm{\oii}=\dfrac{a_Z^2}{2}\dfrac{1}{T^\mathrm{\oiii}}\,,\ a_Z=-12030.22\dfrac{Z}{Z_\odot}+9178.14\,.
\end{equation}\par
The metallicity of a simulated galaxy is usually defined as the gas particle mass-weighted metallicity. However, we will use the $Q_\mathrm{\hi}$-weighted metallicity averaged over all \hii\ regions within a galaxy to determine the gas temperature. This is because the direct $T_e$ method constrains the \oiii\ and \oii\ ion abundances within a galaxy through the $L_{32}^\mathrm{\oiii}/L_\mathrm{H\beta}$ and $L_{10,20}^\mathrm{\oiii}/L_\mathrm{H\beta}$ measurements. When the \hii\ region gas densities are much lower than the critical densities of the optical [\oiii], [\oii], and H$\beta$ lines, Eq.~(\ref{eq:model_lownH}) shows that $L_{32}^\mathrm{\oiii}/L_\mathrm{H\beta}$ is proportional to
\begin{equation}\label{eq:Z_R3}    Z_\mathrm{R3}=\dfrac{\Sigma ZQ_\mathrm{\hi}V_\mathrm{\oiii}/V_\mathrm{\hii}}{\Sigma Q_\mathrm{\hi}}\,,
\end{equation}
 and $L_{10,20}^\mathrm{\oiii}/L_\mathrm{H\beta}$ is proportional to
\begin{equation}\label{eq:Z_R2}
    Z_\mathrm{R2}=\dfrac{\Sigma ZQ_\mathrm{\hi}V_\mathrm{\oii}/V_\mathrm{\hii}}{\Sigma Q_\mathrm{\hi}}\,.
\end{equation}
Here the sum runs over all of the \hii\ regions within the target galaxy, each with a unique $Q_\mathrm{\hi}$, $Z$, and $n_\mathrm{H}$. Eqs.~(\ref{eq:Z_R3})--(\ref{eq:Z_R2}) are therefore the
\oiii\ and \oii\ ionic abundances which are  constrained in the
direct $T_e$ method in the low gas density limit. 
%therefore the definitions of \oiii\ and \oii\ ion abundances constrained by the direct $T_e$ method under the low gas density limit. 
Notice that throughout this work we assume that an \hii\ region is occupied by either \oiii\ or \oii, and so the volume correction factors cancel out when combining \oiii\ and \oii\ ionic abundances: 
\begin{equation}
Z_\mathrm{Q}=Z_\mathrm{R3}+Z_\mathrm{R2}=\dfrac{\Sigma ZQ_\mathrm{\hi}}{\Sigma Q_\mathrm{\hi}}\,.
\end{equation}
It is therefore apparent from Eq.~(\ref{eq:model_lownH}) that the metallicity measured by the direct $T_e$ method is the $Q_\mathrm{\hi}$-weighted metallicity averaged over all \hii\ regions within a galaxy. Note that the $T_4^\mathrm{\oii}$ model fits from \cite{2020A&A...634A.107Y} predict that 3 out of 22 \textsc{FIRE} galaxies have $T^\mathrm{\oii}<5000$\,K. We however adopt a floor \oii\ region temperature of 5000\,K since still lower temperatures are not observed in
\hii\ regions
%no local galaxy has been observed with such a low temperature 
\citep[e.g.][]{2006A&A...448..955I,2009MNRAS.398..485P,2020A&A...634A.107Y}.\par
We do not fully trust the gas temperatures given directly by FIRE for the following reasons: In FIRE, gas particles are photoionized radially outward from their nearby stellar particles, however to conserve photons at all resolutions the algorithm resorts to stochastic ionization. Once a gas particle is randomly ionized, its temperature is simply set to $10^4$\,K. This may be too hot for metal enriched dusty \hii\ regions or too cold for metal-poor photo-heated gas, or for gas particles facing harder ionizing spectra. As a result, the \hii\ region gas temperatures in FIRE are close to $T_4 = 1$ by construction \citep{2018MNRAS.480..800H,2020MNRAS.491.3702H}, which is low compared with current high-$z$ observational constraints \citep{2022arXiv220712375C}. Moreover, our model self-consistently solves for the sub-resolution ionization state of the gas. Therefore, we estimate the \hii\ region gas properties around each stellar particle by averaging over its gas particle neighbors without referring to the randomly-assigned ionization state from FIRE. For example, there may be gas particles located close to a star particle that are not randomly ionized in FIRE and are hence assigned spuriously low temperatures. To rapidly estimate the volume correction factors $V_\mathrm{\oiii}/V_\mathrm{\hii}$ and $V_\mathrm{\oii}/V_\mathrm{\hii}$ for each \hii\ region, we add four finely gridded dimensions covering $-5.0\leq \log n_\mathrm{H}/[\mathrm{cm}^{-3}]\leq4.0$, $-4.0\leq\log Z/[\mathrm{Z}_\odot]\leq0.6$, $40.5\leq\log Q_\mathrm{HI}/[\mathrm{s}^{-1}]\leq51.5$, and $0.5\leq T_4\leq 9.0$ to the existing FSPS lookup table. We then create a 6D volume correction factor lookup table using the model introduced in Section~\ref{subsec:volume}. Thus, for each gas particle we determine its radiation spectrum and ISM environment, and we then assume that each star particle creates a uniform \hii\ region and use the model introduced in Section~\ref{sec:model} to estimate the [\oiii], [\oii], H$\alpha$, and H$\beta$ line luminosities.\par
\par

\begin{figure*}
    \centering
    \includegraphics[width=0.41\textwidth,trim={0 0 3cm 0},clip]{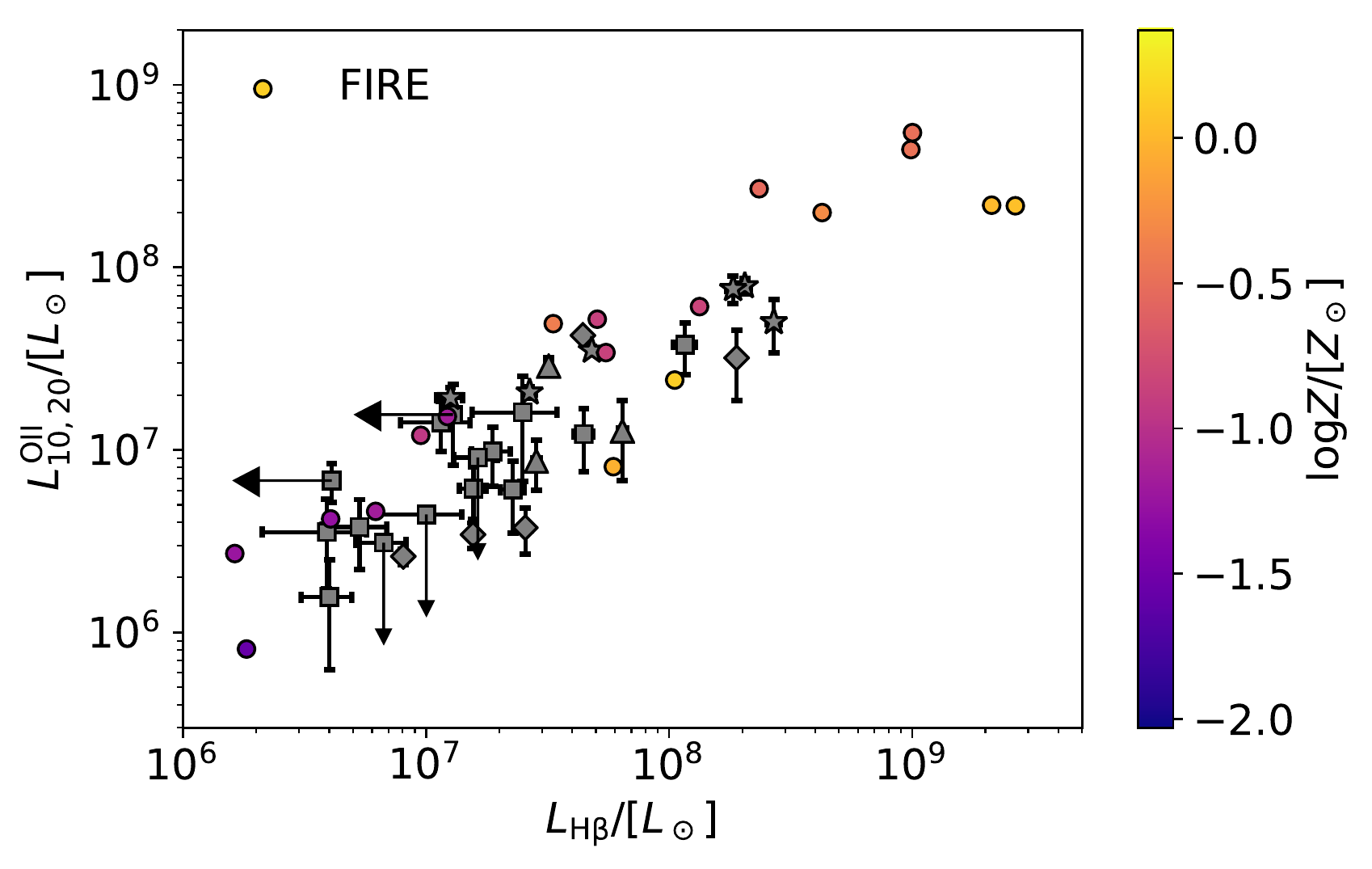}
    \includegraphics[width=0.5\textwidth]{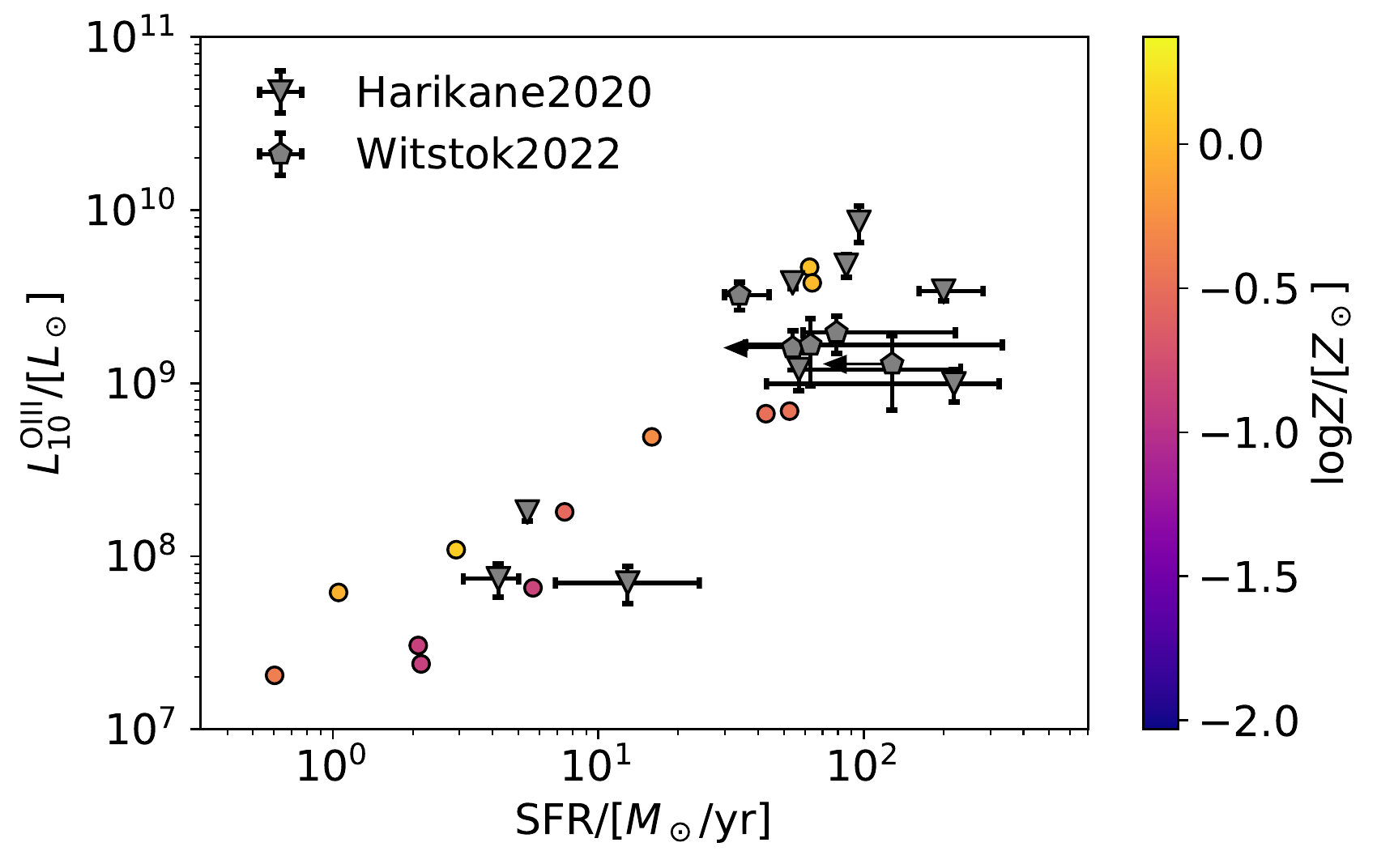}\\
     \includegraphics[width=0.41\textwidth,trim={0 0 3cm 0},clip]{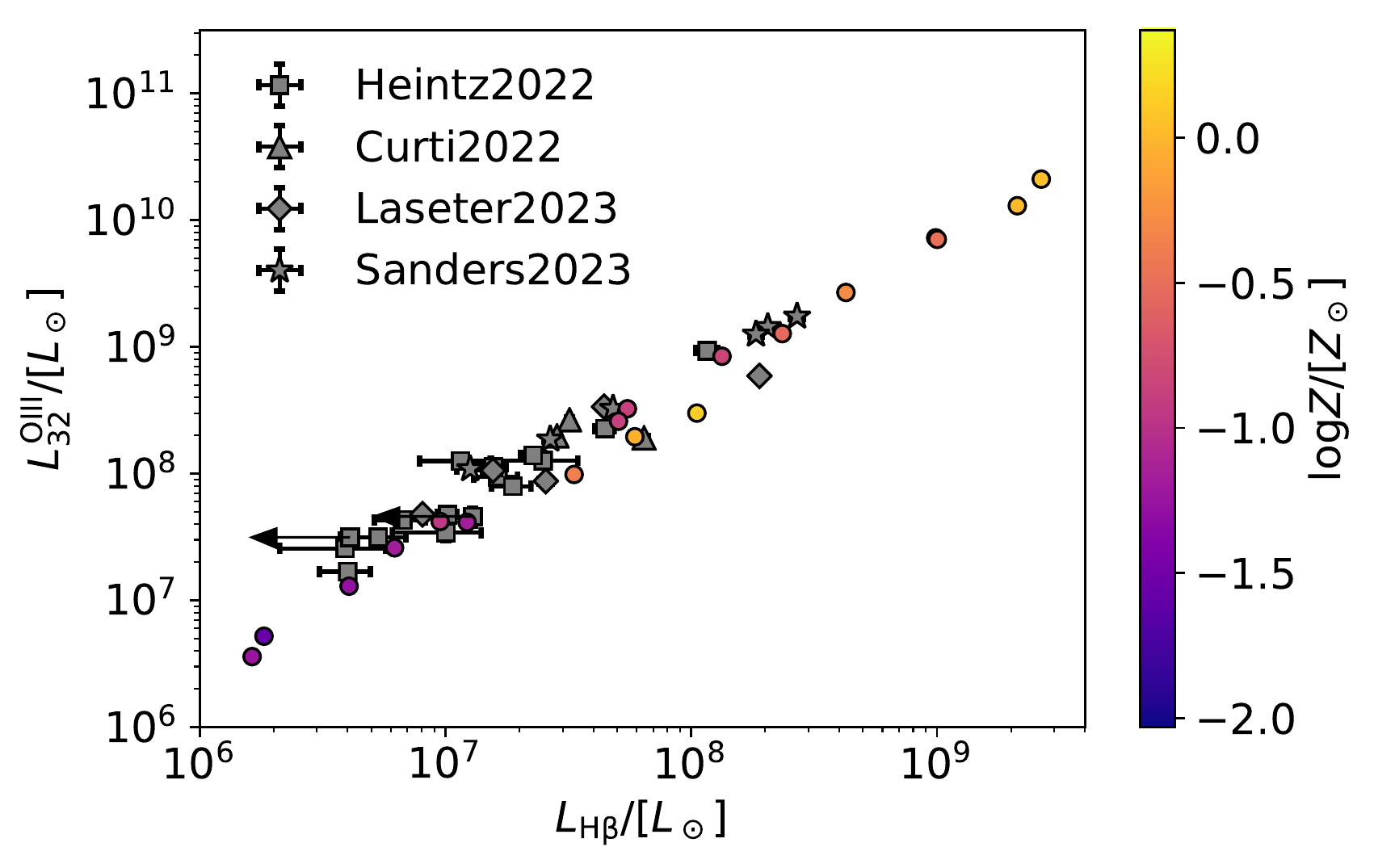}
     \includegraphics[width=0.5\textwidth]{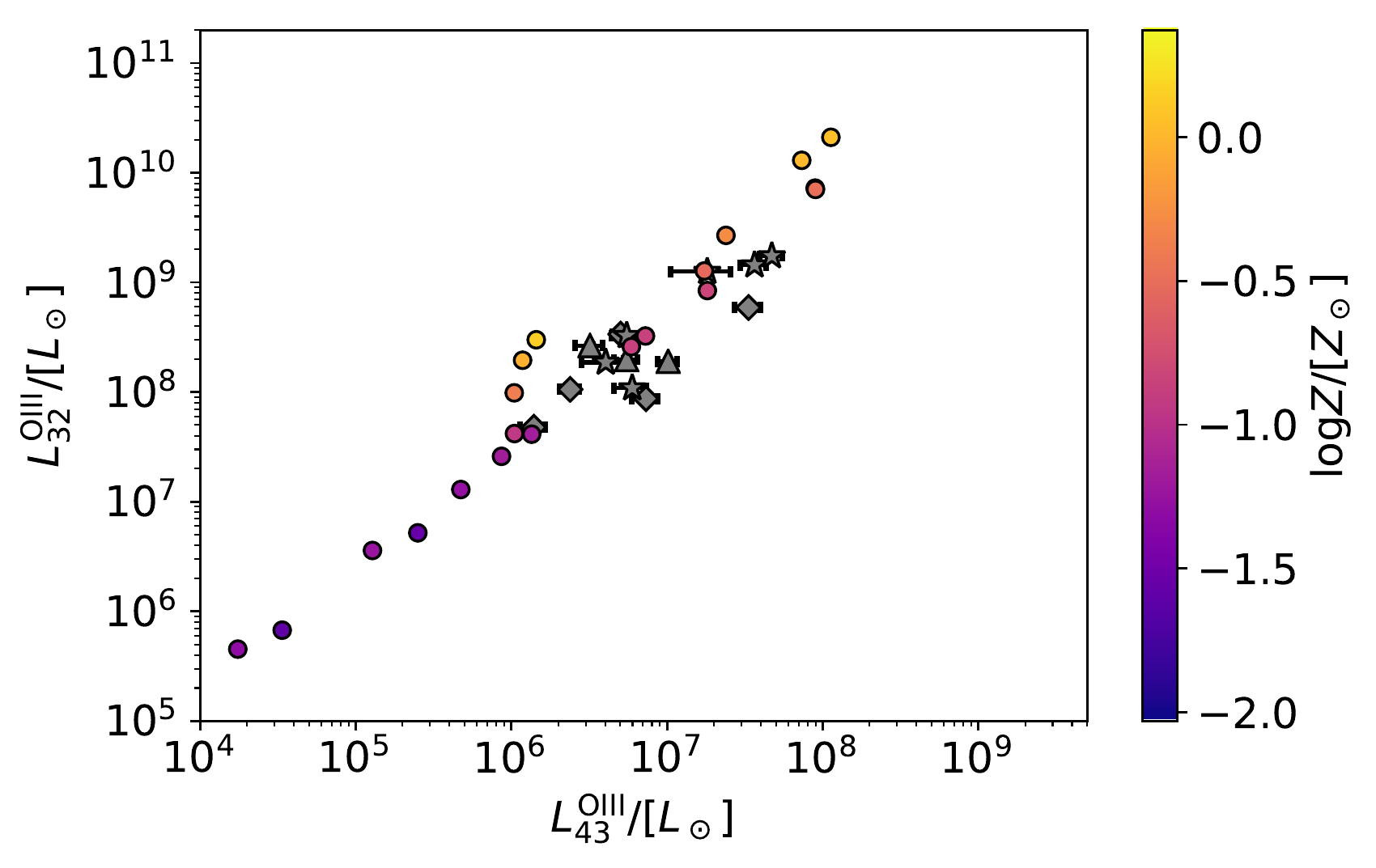}
    \caption{Comparison between the post-processed FIRE ISM emission line luminosities with \textit{JWST} and ALMA measurements for [\oiii] 5008\AA, 4364\AA, 88\,$\mu$m, H$\beta$, and [\oii] 3727,30$\angstrom$lines. In each figure, the points colour coded by $Q_\mathrm{\hi}$-weighted} metallicity show the post-processed line luminosities for the primary FIRE galaxies at $z=6$. The data points are from \protect\cite{2020ApJ...896...93H,2022MNRAS.515.1751W,2022arXiv221202890H,2022arXiv220712375C,2023arXiv230308149S,2023arXiv230603120L} at $6<z<9.5$, as specified by the legends. Our model predictions are in good agreement with observational results.\label{fig:L_JWST}
\end{figure*}

We compare the post-processed FIRE ISM line luminosities with recent \textit{JWST} \citep{2022arXiv220712375C,2022arXiv221202890H,2023arXiv230308149S,2023arXiv230603120L} and ALMA \citep{2020ApJ...896...93H,2022MNRAS.515.1751W} measurements from redshift $6<z<9.5$ in Figure~\ref{fig:L_JWST}. We have corrected the optical line luminosities of \textit{JWST} targets ID4590 and ID10612 from \cite{2022arXiv220712375C}, and all sources in \cite{2023arXiv230308149S,2023arXiv230603120L} for dust attenuation, adopting an extinction curve with $R_\mathrm{V}=2.5$ \citep{2022arXiv220712375C}. In the case of the other \textit{JWST} galaxies, the dust attenuation, $A_\mathrm{V}$, is measured to be smaller than $0.25$ magnitudes (and mostly less than 0.1 magnitudes), and
so dust corrections should be negligibly small. 
The upper left panel presents the [\oii] 3727,30$\angstrom$versus H$\beta$ line luminosity relation, which is sensitive to the gas-phase metallicity, temperature and incident spectrum of the \hii region. The [\oiii] 88\,$\mu$m versus SFR relation, which is sensitive to gas density and metallicity, is presented in the upper right panel. The [\oiii] 5008$\angstrom$versus H$\beta$ line luminosity relation presented in the bottom left panel is a good diagnostic for the average gas phase metallicity in the HII regions. 
Finally, the [\oiii] 5008$\angstrom$versus [\oiii] 4364$\angstrom$line luminosity relation shown in the bottom right panel is a sensitive \oiii\ region gas temperature diagnostic. The FIRE simulation post-processing results are shown as the colour-coded points, where the colour code shows the galaxy metallicities. The more luminous galaxies tend to be more metal-rich and have larger stellar masses. In general, our model predictions are in good agreement with observational results.\par
 
\par

\begin{figure}
    \centering
    \includegraphics[width=0.5\textwidth]{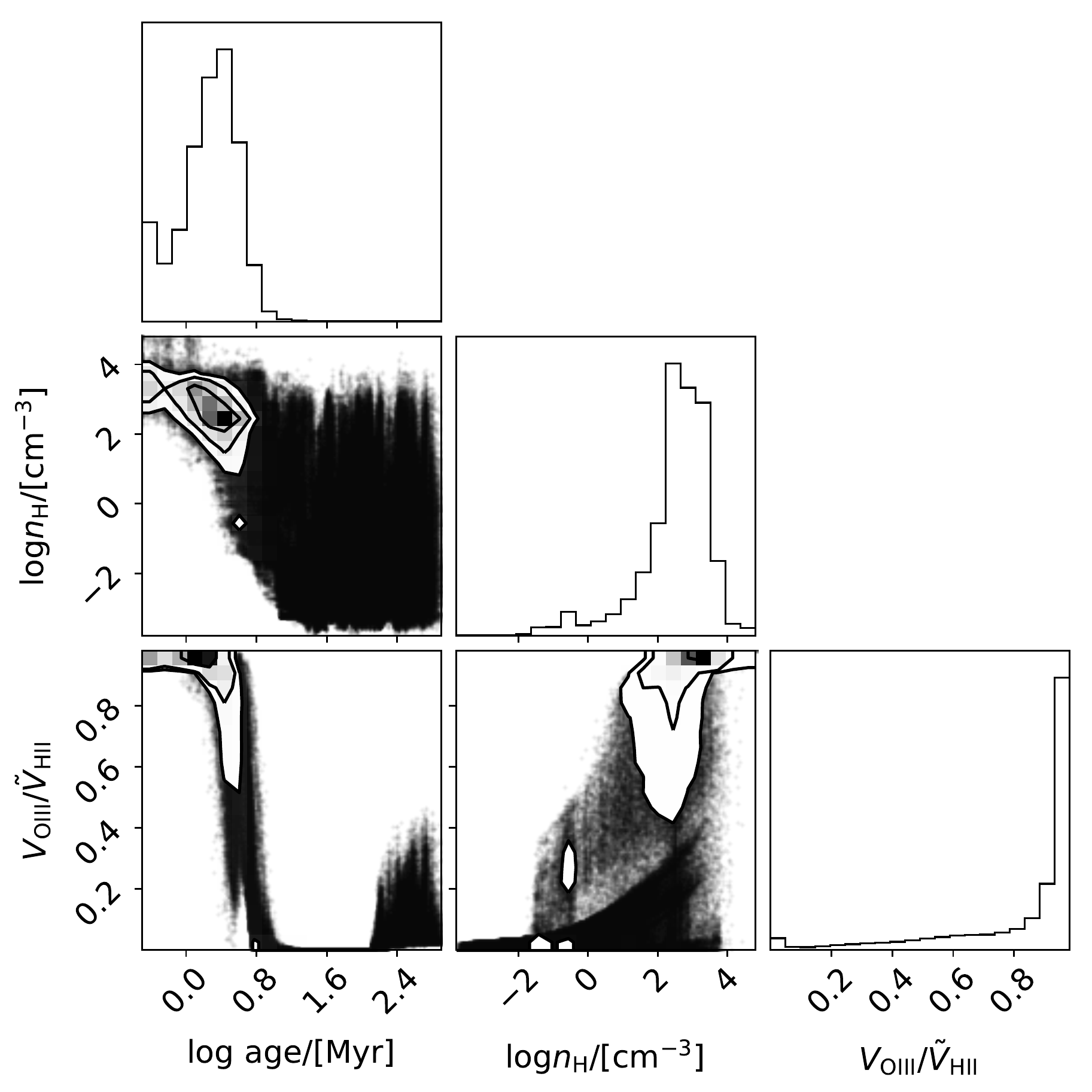}
    \caption{Distributions of $Q_\mathrm{\hi}$, the hydrogen number density $n_\mathrm{H}$ within each \hii\ region, and 
    the volume correction factor for \oiii\ around
     all of the stellar particles in the FIRE high-$z$ galaxy z5m12b at $z=6$. All histograms are weighted by the value of $L_\mathrm{H\beta}$ for each individual \hii\ region. Young and massive stellar particles with high hydrogen ionizing photon production rates, $Q_\mathrm{\hi}$, tend to live in denser gas environments.}\label{fig:ISM2Dhist}
\end{figure}

Figure~\ref{fig:ISM2Dhist} shows the distributions of age, hydrogen number density, $\log n_\mathrm{H}$, and the \oiii\ region volume correction factor for all stellar particles and their nearby \hii\ regions within the most massive primary galaxy z5m12b. All histograms are weighted by the \hii\ region H$\beta$ line luminosity. The majority of stellar particles are older than 10 Myr, corresponding to $\log Q_\mathrm{HI}/[\mathrm{s}^{-1}]\lesssim49$. With such soft and weak incident spectra, the \hii regions sourced from these older stellar populations are dominated by \oii. Only 6.5 per cent of the stellar particles shown are younger than 10 Myr, and these particles make up just 7.9 per cent of the total stellar mass. However, these young stellar populations contribute 99.6 per cent of the total galaxy-wide $Q_\mathrm{\hi}$. Star particles with more intense UV radiation tend to live within denser gas environments, as shown by the correlation in the $\log n_\mathrm{H}$--$\log\ \mathrm{age}$ 2D distribution at $\mathrm{age}\lesssim 10$ Myr.\par

\par

\begin{figure}
    \centering
    \includegraphics[width=0.5\textwidth]{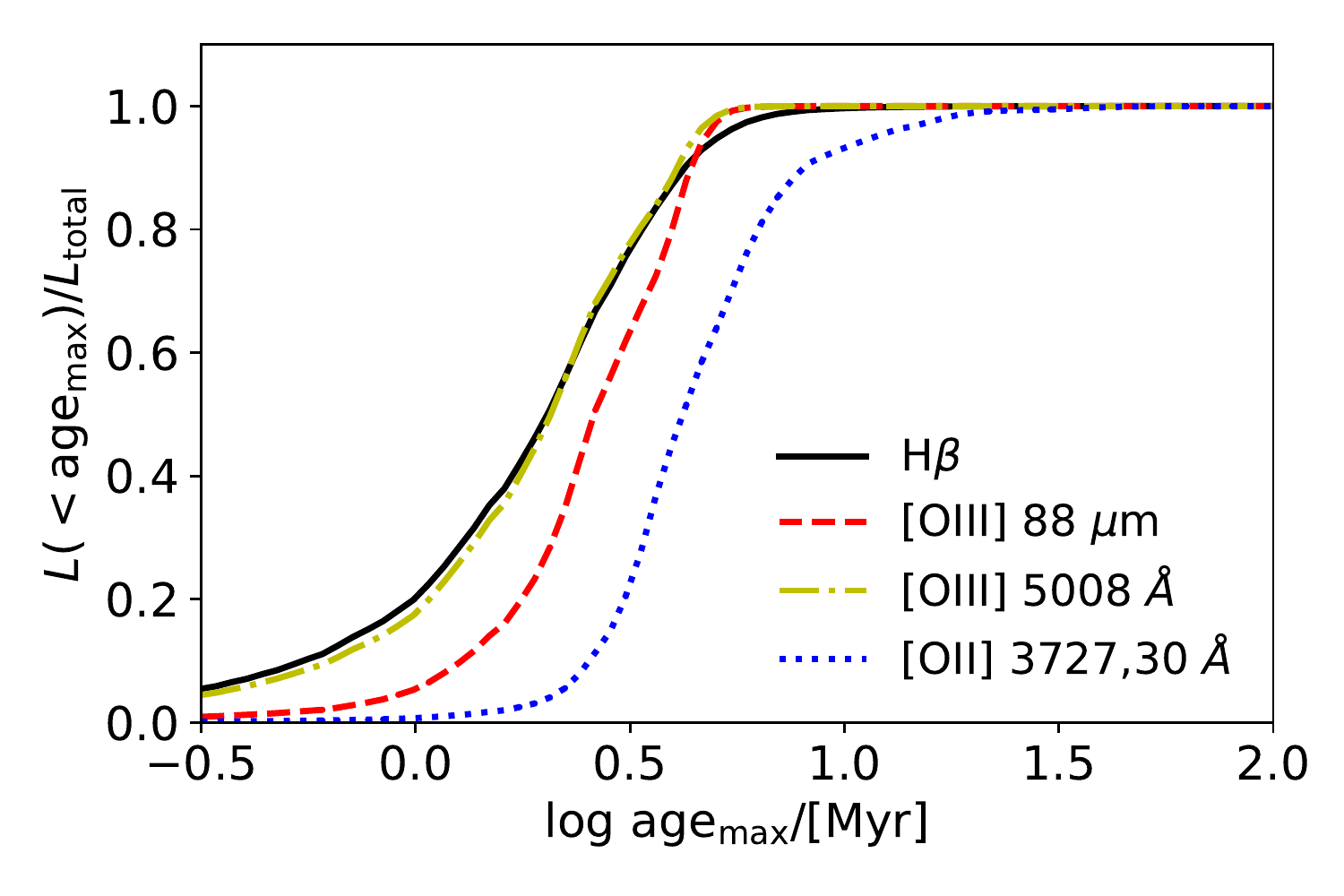}
    \caption{Fractional line luminosities $L(\mathrm{age}\leq\mathrm{age_{max}})/L_\mathrm{tot}$ for FIRE galaxy z5m12b. [\oiii] and H$\beta$ lines are mainly sourced by stellar populations younger than 6 Myr, while [\oii] lines trace slightly older stellar populations ($\mathrm{age}\lesssim10$ Myr) with softer radiation spectra.}\label{fig:LQ}
\end{figure}

Figure~\ref{fig:LQ} further shows the fractional contributions to the [\oiii] 88\,$\mu$m, 5008\AA, [\oii] 3727,30\AA, and H$\beta$ line luminosities
from \hii\ regions surrounding stellar particles in the simulated FIRE galaxy z5m12b, as a function of stellar ages. We find that the [\oiii] and H$\beta$ lines are mostly contributed by stellar populations younger than 6 Myr, while a significant fraction of the [\oii] line luminosity is contributed by slightly older stellar populations (6 Myr - 10 Myr). This can be understood through the \oii\ fractional abundance radial distribution shown in Figure~\ref{fig:OIIIProfile} and the $V_\mathrm{\oiii}/\tilde{V}_\mathrm{\hii}$ distribution presented in Figure~\ref{fig:ISM2Dhist}. Although young stellar populations with higher $Q_\mathrm{\hi}$ tend to source larger \hii\ regions, the harder incident spectrum from such populations tends to doubly ionize surrounding oxygen, and so most of the oxygen near the youngest stars is in \oiii\ rather than \oii. As a result, stellar populations older than 6 Myr contribute about 24 per cent of the [\oii] luminosity in this galaxy, whereas this fraction is only 0.2 per cent for the [\oiii] lines. We find that this trend, in which [\oii] lines trace slightly older stellar populations than the [\oiii] lines, holds in all FIRE high-$z$ suite primary galaxies.\par
\begin{figure*}
    \centering
    \includegraphics[width=0.45\textwidth]{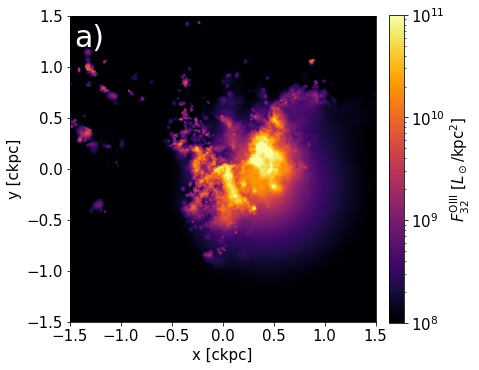}
    \includegraphics[width=0.45\textwidth]{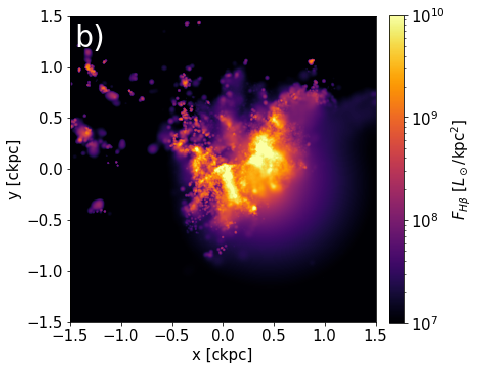}\\
    \includegraphics[width=0.45\textwidth]{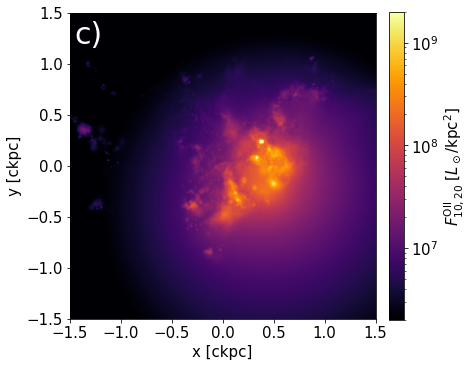}
    \includegraphics[width=0.45\textwidth]{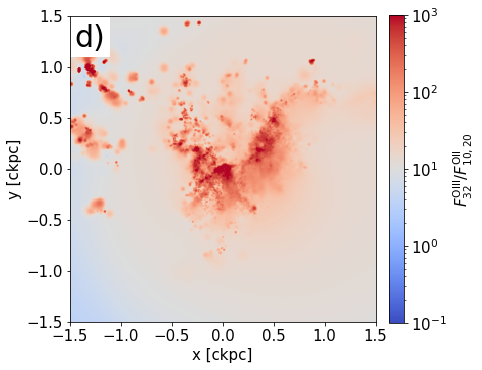}
    \caption{The line surface brightness distributions of the FIRE galaxy z5m12b at $z=6$. Panel a), b), c) show [\oiii] 5008\AA, H$\beta$, and [\oii] 3727,30$\angstrom$surface brightnesses respectively. Panel d) shows the surface brightness ratio between [\oiii] 5008$\angstrom$and [\oiii] 3727,30\AA, which traces stellar populations younger than a few Myr.}\label{fig:Flux}
\end{figure*}

 
To better illustrate this, we show the [\oiii] 5008\AA, H$\beta$, and [\oii] 3727,30$\angstrom$line surface brightness distributions for z5m12b in Figure~\ref{fig:Flux} a), b), c). 
These show face-on images of the simulated galaxy, zoomed in to the central region
where the stellar populations younger than $\sim10$ Myr are mostly concentrated. The surface brightness distributions [\oiii], H$\beta$, and [\oii] are not identical because [\oiii] lines trace the product of $Q_\mathrm{\hi}$ and the metallicity, $Z$, of the \hii\ region around each star particle, while the $H\beta$ line depends only on the $Q_\mathrm{\hi}$ distribution, and the [\oii] lines trace the product of $Q_\mathrm{\hi}$, $Z$, and the volume correction factor, $V_\mathrm{\oii}/V_\mathrm{\hii}$.
We present the ratio between the [\oiii] and [\oii] surface brightness in panel d). The 
2D-pixels with $F^\mathrm{\oiii}_{32}/F^\mathrm{\oii}_{10,20}\gtrsim100$ trace very young stellar populations of age less than a few Myr.\par
\par

\par

\section{ISM inhomogeneity versus the one-zone approximation}\label{sec:inhomo}
The one-zone approximation is widely adopted to determine the gas properties from ISM emission line measurements. Specifically, a galaxy's ISM is often treated as a uniform sphere or as an infinite slab of gas. This is clearly a highly simplified treatment as, in reality, the ISM properties vary
across a galaxy, which itself has a complex, multi-component geometry. 
In addition, the stellar populations responsible for ionizing the surrounding gas may have a broad range of ages and diverse spectral shapes. That is, the one-zone calculations neglect inhomogeneities in the ISM properties and adopt simplified descriptions of the stellar populations. 
Therefore, the physical meaning of the ISM parameter inferences from one-zone models
is somewhat unclear.\par 
The FIRE high-$z$ suite with post-processed ISM emission provides a powerful tool to test this one-zone assumption. In this section, we combine the simulated galaxy-wide ISM line luminosities with the one-zone description to constrain the characteristic ISM properties for each FIRE primary galaxy at $z=6$.
This allows us to test how the parameters inferred from a one-zone model description of the FIRE galaxies compare with their actual properties. We will test the one-zone model under an ideal scenario where multiple [\oiii], [\oii], and H$\beta$ lines are resolved by \textit{JWST} for high-z galaxies in Section~\ref{subsec:multiline}. The case where [\oiii] 88\,$\mu$m and 52\,$\mu$m lines are both resolved by ALMA will be discussed in Section~\ref{subsec:multisubmmline}. Finally we discuss the more practical ALMA case where only the [\oiii] 88\,$\mu$m line is detected in Section~\ref{subsec:singleline}.\par
As discussed in Section~\ref{sec:FIRE}, we have assumed that the gas temperature for \hii, \oiii, and \oii\ regions are constant throughout each galaxy when post-processing the FIRE ISM line emission. Therefore, we account for inhomogeneity only in the gas density, metallicity, and incident spectrum, but ignore the non-uniform temperature distribution, which may cause biases in metallicity constraints \citep[e.g.][]{2005A&A...434..507S,2022arXiv221014234C}.\par

\subsection{One-zone ISM constraints with multiple optical line detections}\label{subsec:multiline}
In this sub-section we combine the simulated galaxy-wide [\oiii], [\oii], and H$\beta$ line luminosities with the analytical model introduced in Section~\ref{sec:model} to solve for the characteristic \hii\ region gas density $n_\mathrm{H}$ and gas phase metallicity $Z$, assuming the galaxy is one-zone. We consider an ideal scenario for the \textit{JWST} high-z measurements, where $L_{32}^\mathrm{\oiii}$, $L^\mathrm{\oiii}_{43}$, $L^\mathrm{\oii}_{10,20}$, and $L_\mathrm{H\beta}$ are detected. In this case, the \oiii\ region temperature $T_4^\mathrm{\oiii}$ is constrained from the $L^\mathrm{\oiii}_{43}/L_{32}^\mathrm{\oiii}$ ratio. The galaxy-wide $Q_\mathrm{\hi}$ is also known from the H$\beta$ line luminosity. The unknown ISM parameters are the gas phase metallicity, gas density, and the \oii\ region temperature. When post-processing \textsc{FIRE} emissions we have assumed a `true' $T_4^\mathrm{\oii}(T_4^\mathrm{\oiii})$ relation as given by \cite{2020A&A...634A.107Y}. Here we assume the $T_4^\mathrm{\oii}(T_4^\mathrm{\oiii})$ relation fit by \cite{2006A&A...448..955I} for the one-zone ISM parameter interpretations to mimic the effects of errors in the \oii\ region temperature estimates. We assume a typical gas density of $n_\mathrm{H}=100$ cm$^{-3}$ in the one-zone metallicity constraints.\par 
One important application for ISM metallicity measurements is to constrain the stellar mass--metallicity relation (MZR). In lower-redshift galaxy samples, there is a well-established correlation between the gas phase metallicity and stellar mass \citep{1979A&A....80..155L,2004ApJ...613..898T}. The MZR is thought to reflect, in part, the impact of outflows, which drive gas and metals out of the shallow potential wells of low-mass galaxies, but have less effect in larger galaxies. As such, the shape and normalization of this relationship and its redshift evolution provide crucial input for models of galaxy formation and evolution, and regarding the feedback processes that regulate galaxy growth.\par
\begin{figure}
    \centering
    \includegraphics[width=0.5\textwidth]{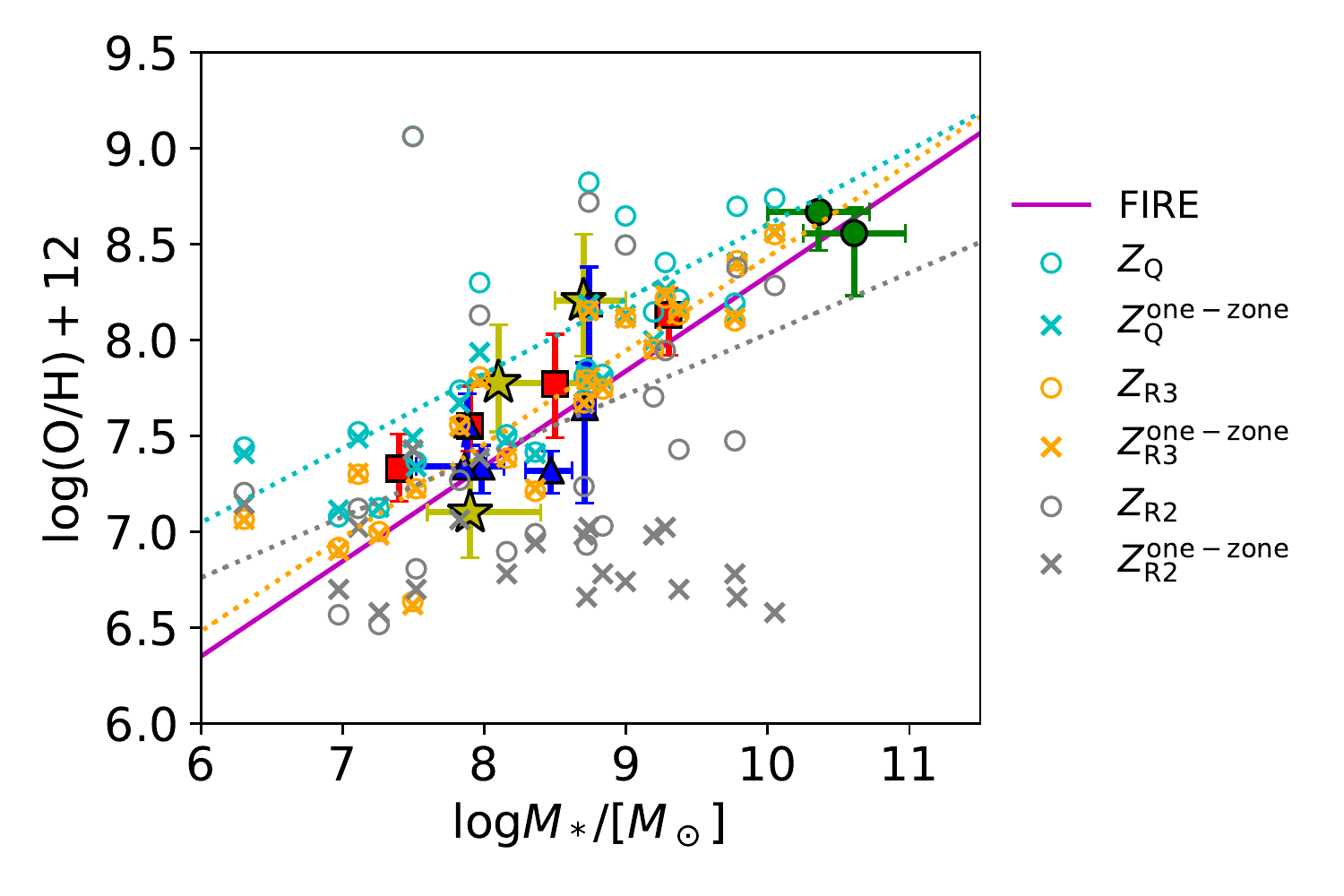}
    \caption{The MZR comparison between ALMA \citep[green;][]{2020ApJ...896...93H,2020MNRAS.499.3417Y} at $z\sim6$; \textit{JWST} metallicity measurements at $z=5$--$7$ red, \citep[red;][]{2022arXiv221108255M} and $z=7$--$10$ \citep[yellow;][]{2022arXiv220712375C}; \citep[blue;][]{2022arXiv221202890H}; FIRE galaxy mass-weighted metallicities averaged over all gas particles \citep[magenta line;][]{2016MNRAS.456.2140M}; $Q_\mathrm{\hi}$-weighted metallicities averaged over all stellar particles (cyan points); $Z_\mathrm{R3}$ and $Z_\mathrm{R2}$ defined in Eq.~(\ref{eq:Z_R3})--(\ref{eq:Z_R2}) (orange and grey circles); \oiii, \oii, oxygen abundance solutions for the one-zone model (orange, grey, and cyan crosses). We present the linear best-fit MZR for $Z_\mathrm{Q}$, $Z_\mathrm{R3}$, and $Z_\mathrm{R2}$ as cyan, orange, and grey dotted lines, respectively. The FIRE MZRs of different definitions at $z=6$ are consistent with the ALMA and \textit{JWST} metallicity measurements. The one-zone metallicity constraints derived from mock observations are close to the mass-weighted metallicity over all gas particles in an inhomogeneous ISM environment.}\label{fig:MZR} 
\end{figure}

 
In the top panel of Figure~\ref{fig:MZR} we compare the FIRE MZR at $z=6$ with ALMA and \textit{JWST} measurements in the redshift range $z=5$--$10$. Specifically, metallicity constraints for two ALMA [\oiii] targets J0217 and J1211 are presented as green circles and crosses \citep{2020ApJ...896...93H,2020MNRAS.499.3417Y}; metallicity constraints for 133 \textit{JWST} [\oiii] emitters at $z=5$--$7$ are given as red squares \citep{2022arXiv221108255M}. Metallicity measurements for nine \textit{JWST} [\oiii] targets at $z=7$--$10$ are shown as yellow stars \citep{2022arXiv220712375C} and blue squares \citep{2022arXiv221202890H}. The one-zone metallicities $Z_\mathrm{R3}^\mathrm{one-zone}$ and $Z_\mathrm{R2}^\mathrm{one-zone}$ are defined as the values that reproduce the galaxy-wide $L^\mathrm{\oiii}_{32}/L_\mathrm{H\beta}$ and $L^\mathrm{\oii}_{10,20}/L_\mathrm{H\beta}$ ratios. Volume correction factors are absorbed into the one-zone metallicities and are therefore ignored in the theoretical model. We present $Z_\mathrm{R3}^\mathrm{one-zone}$ and $Z_\mathrm{R2}^\mathrm{one-zone}$ for each \textsc{FIRE} galaxy as orange and grey crosses in Figure~\ref{fig:MZR}, and $Z_\mathrm{R3}$ and $Z_\mathrm{R2}$, defined in Eqs.~(\ref{eq:Z_R3})--(\ref{eq:Z_R2}), as orange and grey circles. As expected, $Z_\mathrm{R3}^\mathrm{one-zone}$ is identical to $Z_\mathrm{R3}$ for all galaxies since the \oiii\ region temperatures assumed in the one-zone model are accurate. The excellent agreement between $Z_\mathrm{R3}^\mathrm{one-zone}$ and $Z_\mathrm{R3}$ is independent of the $n_\mathrm{H}$ assumption since the critical densities for [\oiii] 5007\,$\angstrom$ and H$\beta$ lines are very high, making the low density limit an excellent approximation.
We find that the low density assumption is also accurate for interpreting [\oii] line measurements. 
%We find the low density assumption is also excellent for [\oii] line interpretations. 
However, since the $T_4^\mathrm{\oii}$ model fits from \cite{2006A&A...448..955I} generally predict higher \oii\ region temperatures than the `true' values we have assumed in simulations, the one-zone model overestimates the [\oii] line emissivities and further underestimates $Z_\mathrm{R2}$. The total one-zone metallicities $Z_\mathrm{Q}^\mathrm{one-zone}=Z_\mathrm{R3}^\mathrm{one-zone}+Z_\mathrm{R2}^\mathrm{one-zone}$ (cyan crosses) are also generally lower than $Z_\mathrm{Q}$ (cyan circles) due to the $T_4^\mathrm{\oii}$ over-estimates.\par
It is generally inconvenient to compute the value of $Z_\mathrm{Q}$ for a simulated galaxy. $Z_\mathrm{Q}$ depends on the gas-phase metallicity distribution within each galaxy, on the stellar population synthesis model, and on the precise approach
employed in estimating local ISM properties. For example, the $Z_\mathrm{Q}$ of each \textsc{FIRE} galaxy may change if we replace the \textsc{FSPS} stellar radiation spectra with alternate models that account for binaries \citep[e.g.][]{2017PASA...34...58E}. The overall metallicity of simulated galaxies is therefore generally defined by the gas particle mass-weighted metallicity, which is more directly captured in the simulation without additional post-processing. We present the best fit FIRE gas particle mass-weighted metallicity $Z_m$ versus stellar mass relation at $z=6$ as the magenta line in Figure~\ref{fig:MZR}. $Z_\mathrm{Q}$ are higher than $Z_m$ because some metal-poor gas particles contribute to the FIRE MZR estimates but are far from stellar particles, and hence such gas does not produce [\oiii] line emission. In this work, we define the metallicity of an \hii\ region as the mass-weighted metallicity averaged over the 32 nearest gas particle neighbours of each stellar particle. Gas particles that are far away from all the stellar populations are therefore excluded from our metallicity estimates, leading to slightly higher galaxy-wide metallicity.

Overall, $Z_\mathrm{Q}$, $Z_\mathrm{Q}^\mathrm{one-zone}$, and $Z_m$ are generally different by less than a factor of two within the stellar mass range $7\leq\log M_*/[\mathrm{M}_\odot]\leq11$. This is comparable to the metallicity scatter and generally smaller than current measurement uncertainties. The MZR at $z=6$ predicted by the FIRE simulations is consistent with \textit{JWST} and ALMA metallicity measurements. We therefore conclude that, provided that the parameters derived are interpreted appropriately, the one-zone approximation appears to work well for determining the ISM properties from rest-frame optical [\oiii], [\oii], and H$\beta$ emission line observations. It is likely valid to compare the gas-particle mass-weighted metallicities directly with measurements since the $Z_m$ values are only slightly lower than $Z_\mathrm{Q}$, at least assuming the stellar SEDs given by \textsc{FSPS}.\par

\par

\subsection{One-zone ISM constraints with multiple sub-millimeter line detections}\label{subsec:multisubmmline}
In sub-section~\ref{subsec:multiline} we quantitatively verified the one-zone approximation for cases where multiple [\oiii], [\oii] optical lines and the H$\beta$ line luminosities are detected for a galaxy sample. The current ALMA and \textit{JWST} high-$z$ galaxy samples only partially overlap. ALMA 
has made detections of the 
[\oiii] 88\,$\mu$m line from the \hii\ regions in $z\sim6-9$ galaxies. \cite{2021MNRAS.504..723Y} forecast, however, that it is possible for ALMA to detect the [\oiii] 52\,$\mu$m line within reasonable integration times if the typical \hii\ region gas density is higher than $\sim100$ cm$^{-3}$. This information may be combined with SFR estimates based on UV and IR luminosity measurements. The [\oiii] 88\,$\mu$m and 52\,$\mu$m luminosities are insensitive to gas temperature, but do depend on $Q_\mathrm{\hi}$, $n_\mathrm{H}$, and $Z$. In this sub-section we will verify the one-zone model for an ideal scenario where $L^\mathrm{\oiii}_{10}$, $L^\mathrm{\oiii}_{21}$ and SFR measurements are available. We make the typical assumption that the gas temperatures follow $T_4^\mathrm{\hii}=T_4^\mathrm{\oiii}=1$ for the one-zone gas density and metallicity constraints.\par
\begin{figure}
    \centering
    \includegraphics[width=0.5\textwidth]{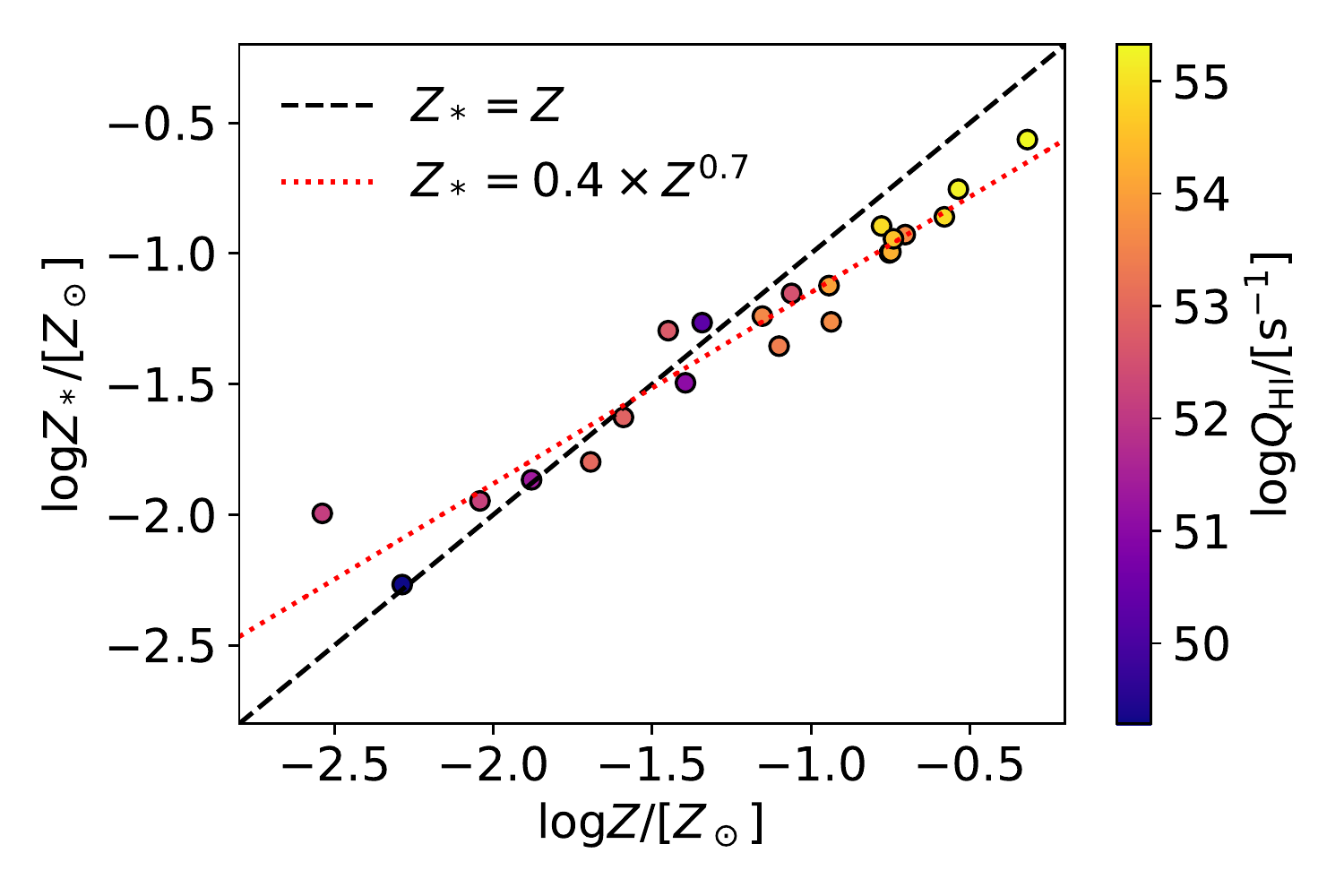}
    \caption{Stellar metallicity, $Z_*$, versus gas-phase metallicity, $Z$, in FIRE. The points specify $Z_*$ and $Z$ for the FIRE high-$z$ galaxies, colour coded by the galaxy-wide $Q_\mathrm{\hi}$. The black dashed line indicates the $Z_*=Z$ case. The red dotted line shows the best-fit $\log Z_*-\log Z$ linear relation. $Z_*\approx Z$ is a good approximation for the FIRE high-$z$ galaxy sample.}\label{fig:ZgasZstar}
\end{figure}
\begin{figure}
    \centering
    \includegraphics[width=0.5\textwidth]{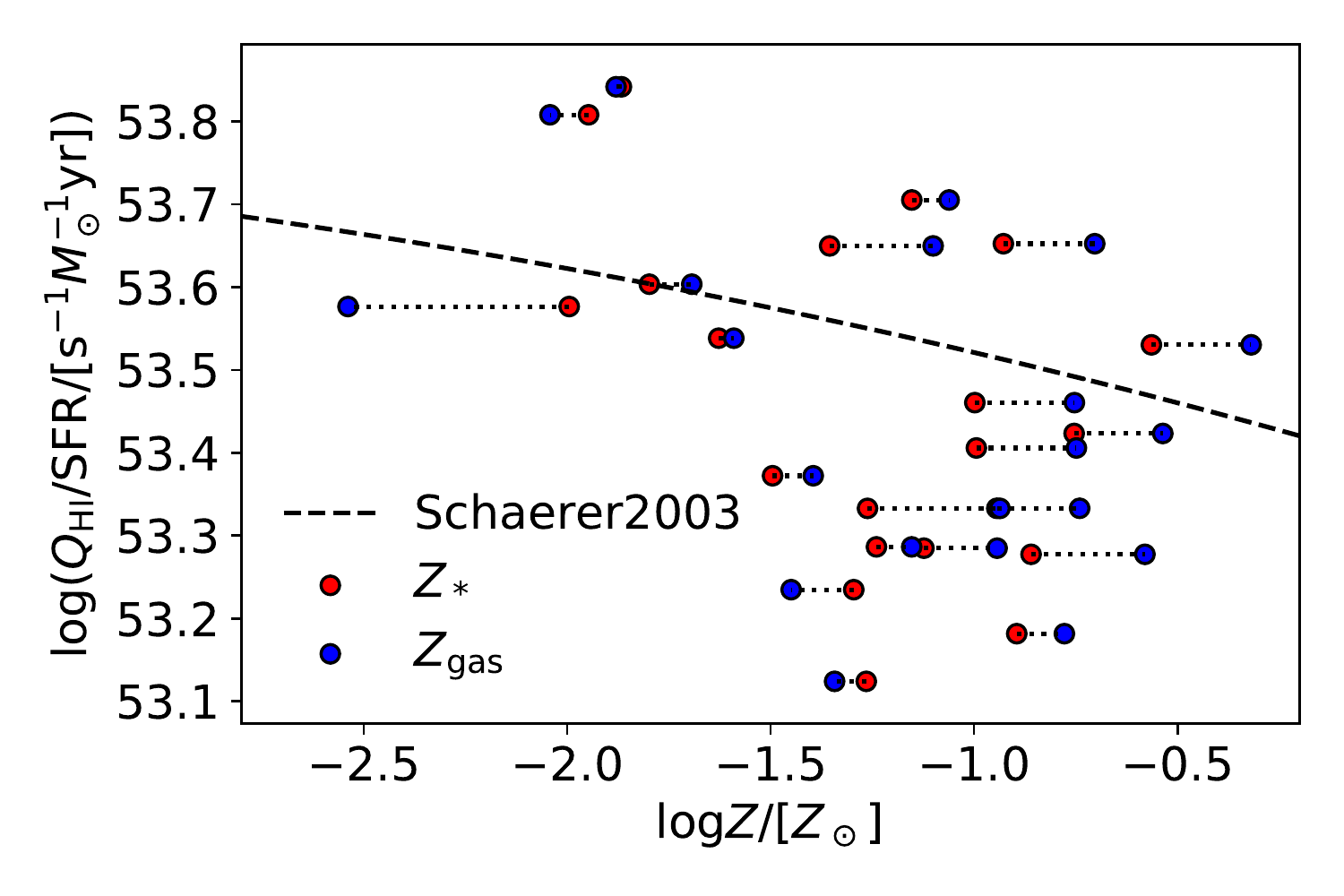}
    \caption{$Q_\mathrm{\hi}/\mathrm{SFR}$ versus metallicity relation. The red and blue point pairs connected by the dotted horizontal lines show the $Q_\mathrm{\hi}/\mathrm{SFR}$ versus average stellar and gas-phase metallicity for each FIRE galaxy. The black dashed line shows the fit from \protect\cite{2003A&A...397..527S} for stellar populations with a Salpeter IMF and an age above 6 Myr, as specified in Eq.~(\ref{eq:Q-SFR}). The \protect\cite{2003A&A...397..527S} relation provides a decent description for the FIRE high-$z$ galaxy sample, although it fails to capture the $Q_\mathrm{\hi}/\mathrm{SFR}$ scatter.}\label{fig:Q2SFR}
\end{figure}

It is useful to connect the instantaneous SFR measurements with the strength of stellar radiation $Q_\mathrm{\hi}$ through stellar population synthesis models. For example, \cite{2003A&A...397..527S} provides a convenient fit for the $Q_\mathrm{\hi}/\mathrm{SFR}$ versus stellar metallicity $Z_*$ for stellar populations older than 6 Myr, assuming a Salpeter IMF \citep{1955ApJ...121..161S}:
\begin{equation}\label{eq:Q-SFR}
\begin{split}
    &\log\left(\dfrac{Q_\mathrm{\hi}/[\mathrm{s}^{-1}]}{\mathrm{SFR}/[M_\odot/\mathrm{yr}]}\right)\\
    =&-0.0029\times\left(\log\left(\dfrac{Z_*}{[Z_\odot]}\right)+7.3\right)^{2.5}+53.81\,.
\end{split}
\end{equation}
To avoid introducing an additional parameter $Z_*$, it is usually assumed that the gas-phase metallicity and stellar metallicity are identical or linearly correlated. In Figure~\ref{fig:ZgasZstar} we show the stellar mass-weighted stellar metallicity $Z_*$ versus gas particle mass-weighted gas phase metallicity among the FIRE high-$z$ primary galaxies. Although the best-fit stellar and gas-phase metallicity relation is $Z_*\approx0.4Z^{0.7}$, the black dashed line in Figure~\ref{fig:ZgasZstar} shows that $Z_*\approx Z$ is also a good approximation among the FIRE mock galaxies. We then test the $Q_\mathrm{\hi}/\mathrm{SFR}-Z_*$ relation Eq.~(\ref{eq:Q-SFR}) among the FIRE galaxies and the results are shown in Figure~\ref{fig:Q2SFR}. Despite the fact that Eq.~(\ref{eq:Q-SFR}) is calibrated with the Salpeter IMF, for which the amplitude at stellar mass $M_*<1M_\odot$ is higher than the Chabrier IMF we have assumed for the stellar particle SED, it nicely captures the trend and overall amplitude of the FIRE galaxy-wide $Q_\mathrm{\hi}/\mathrm{SFR}-Z_*$ relation. This is because Eq.~(\ref{eq:Q-SFR}) is fit for stellar populations of ages above 6\,Myr, while the major $Q_\mathrm{\hi}$ contributors in the FIRE galaxies are stellar populations younger than 6 Myr. The harder average stellar radiation spectra compensate for the lower amplitude of the Chabrier IMF in the small stellar mass range relevant here. However, since each FIRE galaxy contains stellar particles across a wide range in age and mass, Eq.~(\ref{eq:Q-SFR}) fails to capture the factor of $\sim4$ scatter in $Q_\mathrm{\hi}/\mathrm{SFR}$ at a fixed metallicity.\par
\par

\par
\begin{figure}
    \centering
    \includegraphics[width=0.5\textwidth]{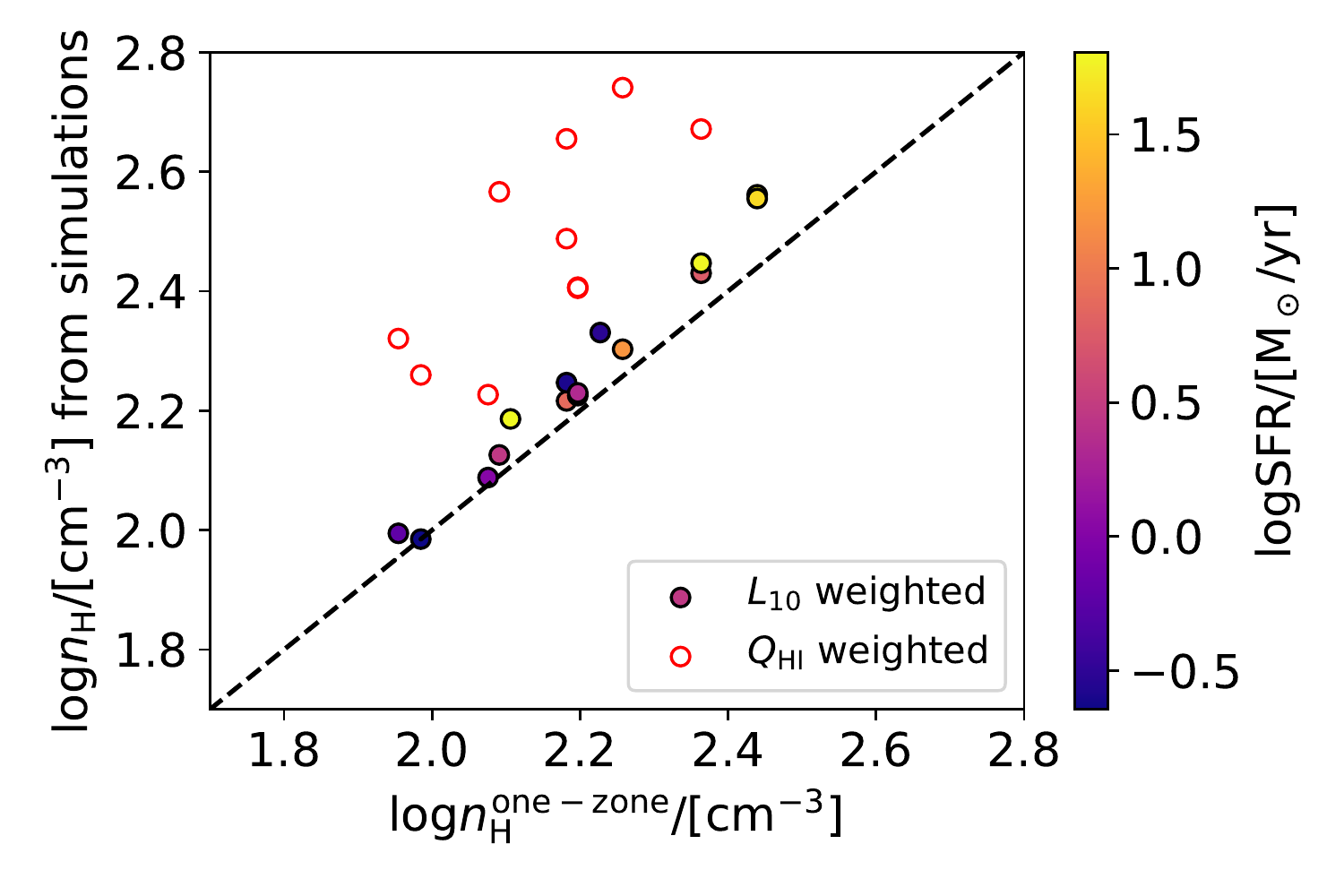}\\
    \includegraphics[width=0.5\textwidth]{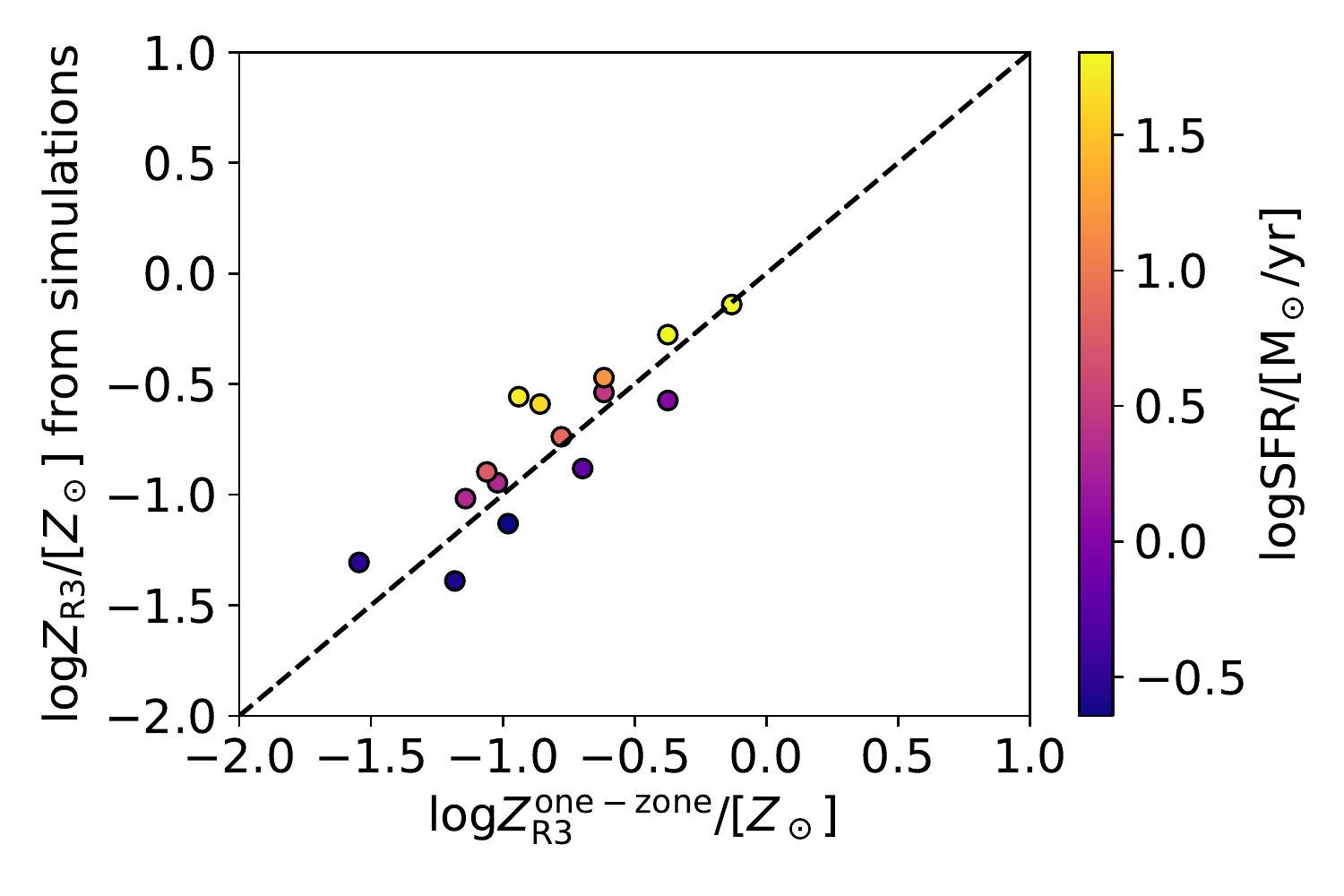}
    \caption{Comparison between the one-zone gas density (top) and metallicity (bottom) solution with simulated ISM properties. Each point corresponds to one FIRE high-$z$ galaxy with SFR$>0.1\ M_\odot/$yr, colour coded by the galaxy-wide SFR. The black dashed line marks the case where ISM properties constrained by one-zone model is identical to simulation results. One-zone gas density provide a good approximation for the [\oiii] 88\,$\mu$m luminosity-weighted gas density, and is comparable to the $Q_\mathrm{\hi}$-weighted gas density (red circles). The one-zone metallicity is close to the $Q_\mathrm{\hi}$-weighted \oiii\ ion abundances, as defined in Eq.~(\ref{eq:Z_R3}).}\label{fig:IR_onezone}
\end{figure}

We define $n_\mathrm{H}^\mathrm{one-zone}$ and $Z_\mathrm{R3}^\mathrm{one-zone}$ as the gas density and metallicity that reproduce the galaxy-wide $L_{10}^\mathrm{\oiii}/\mathrm{SFR}$ and $L_{10}^\mathrm{\oiii}/L_{21}^\mathrm{\oiii}$ for each \textsc{FIRE} galaxy with $\mathrm{SFR}>0.1$ $M_\odot$/yr. Since \oii\ abundances are  inaccessible to ALMA, we here ignore the differences between $Z_\mathrm{Q}$ and $Z_\mathrm{R3}$ in estimating $Q_\mathrm{\hi}$  (Eq.~\ref{eq:Q-SFR}). That is, we assume $V_\mathrm{\oiii}/V_\mathrm{\hii}=1$. We find the one-zone gas densities constrained by the [\oiii] sub-millimeter lines are about three orders of magnitude higher than the gas density averaged over all gas particles. This is because the [\oiii] lines are tracing star-forming regions where the gas densities are much higher than the volume average. In the top panel of Figure~\ref{fig:IR_onezone} we show that $n_\mathrm{H}^\mathrm{one-zone}$ is comparable to the $Q_\mathrm{\hi}$-weighted gas density averaged over all \hii\ regions, and almost identical to the [\oiii] 88 $\mu$m-weighted gas density. This result suggests that direct comparisons between simulated and observed \oiii\ region gas densities are difficult without ISM emission line post-processing analyses. The bottom panel of Figure~\ref{fig:IR_onezone} shows that the values of $Z_\mathrm{R3}^\mathrm{one-zone}$ constrained by [\oiii] sub-millimeter lines are similar to the $Z_\mathrm{R3}$ results given by simulations. The values of $Z_\mathrm{R3}^\mathrm{one-zone}$ and $Z_\mathrm{R3}$ are not identical here because the assumed \oiii\ region gas temperature $T_4^\mathrm{\oiii}=1$ and $Q_\mathrm{\hi}$ given by Eq.~\ref{eq:Q-SFR} are different from the exact values for each galaxy.\par

\subsection{One-zone ISM constraints with single sub-millimeter line detection}\label{subsec:singleline}
In this sub-section, we will verify the one-zone model for more practical cases where only $L^\mathrm{\oiii}_{10}$ and SFR measurements are available. We will still assume $T_4^\mathrm{\hii}=T_4^\mathrm{\oiii}=1$ for the one-zone interpretations.\par  

\begin{figure}
    \centering
    \includegraphics[width=0.5\textwidth]{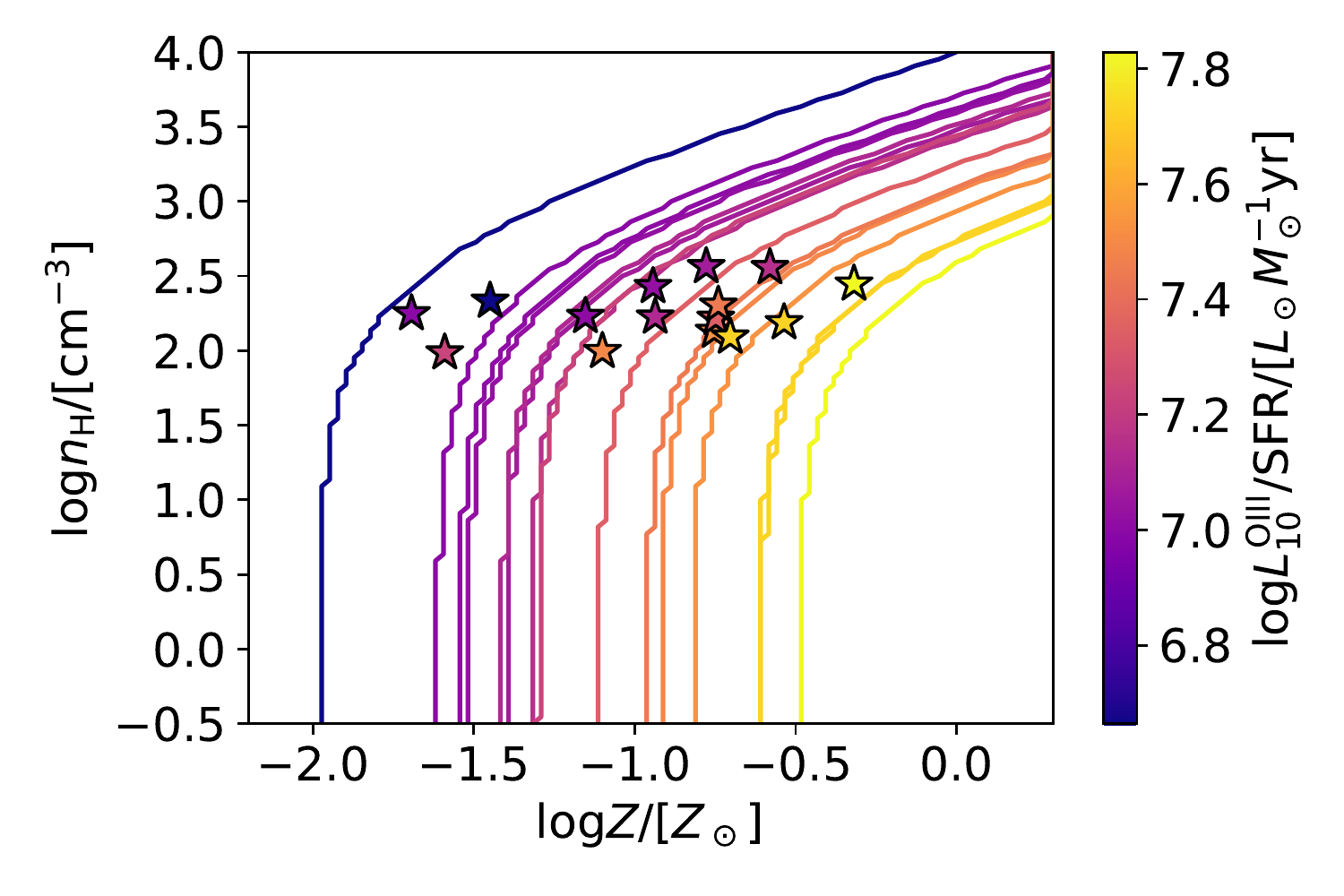}
    \caption{Gas metallicity and density constraints from $L_{10}^\mathrm{\oiii}/\mathrm{SFR}$ and one-zone models for all FIRE galaxies with $\mathrm{SFR}>0.1M_\odot/$yr. The $L_{10}^\mathrm{\oiii}$-weighted gas density and gas particle mass-weighted metallicity for each FIRE galaxy are presented in stars. Differences between the one-zone constraints and the true ISM properties are less than 0.5 dex, which is small compared to ALMA $L_{10}^\mathrm{\oiii}/\mathrm{SFR}$ measurement uncertainties.}\label{fig:nHZ}
\end{figure}

Given the galaxy-wide $L_{10}^\mathrm{\oiii}/\mathrm{SFR}$ for each FIRE galaxy and  Eq.~(\ref{eq:Q-SFR}), one can derive constraints in the
\hii\ region gas density-metallicity plane assuming a one-zone model. We present the $n_\mathrm{H}-Z$ parameter combinations which reproduce the $L_{10}^\mathrm{\oiii}/\mathrm{SFR}$ ratios for all FIRE galaxies with SFR$>0.1 M_\odot/$yr in Figure~\ref{fig:nHZ}. In the low gas density limit $n_\mathrm{H}\ll n_\mathrm{crit,1}^\mathrm{\oiii}=1700$\,cm$^{-3}$, the $L_{10}^\mathrm{\oiii}/Q_\mathrm{\hi}$ given by our model reduces to the simple expression of Eq.~(\ref{eq:model_lownH}). As discussed in Section~\ref{subsec:multisubmmline}, we do not here distinguish between $Z_\mathrm{Q}$ and $Z_\mathrm{R3}$ due to a lack of \oii\ abundance measurements. In this case we have $L_{10}^\mathrm{\oiii}/\mathrm{SFR}\propto L_{10}^\mathrm{\oiii}/Q_\mathrm{\hi}$ independent of gas density. This is why the $n_\mathrm{H}-Z$ constraints become independent of $n_\mathrm{H}$ at low gas densities. In the high gas density limit $n_\mathrm{H}\gg 1700$\,cm$^{-3}$ the \oiii\ ion level population abundances follow
a Boltzmann factor and depend only on temperature. It is easy to show from Eq.~(\ref{eq:LOIIIOII}) that $L_{10}^\mathrm{\oiii}/Q_\mathrm{\hi}\propto Z/n_\mathrm{H}$. Therefore, the $n_\mathrm{H}-Z$ constraints are linearly correlated at high gas densities. Overall the $n_\mathrm{H}-Z$ constraints show an ``L''-shaped degeneracy because two free parameters are being constrained from a single observable. At a fixed $Q_\mathrm{\hi}$, one-zone environments of lower density or higher metallicity tend to be more [\oiii] luminous. This is because in these cases the \oiii\ ions are more abundant, and collisional de-excitations are less effective in competing with the spontaneous decay process, during which the [\oiii] photons are emitted. As a result, galaxies of higher $L_{10}^\mathrm{\oiii}/\mathrm{SFR}$ prefer the $n_\mathrm{H}-Z$ parameter space in the lower right corner of the figure.
In Figure~\ref{fig:nHZ} we also present the galaxy-wide average gas density ($L_{10}^\mathrm{\oiii}$-weighted) and metallicity (gas particle mass-weighted) of FIRE galaxies (star symbols). We find that the $n_\mathrm{H}-Z$ parameter constraints under the one-zone approximation are in general close to the true ISM properties, with a difference less than 0.5 dex. Discrepancies at this level are usually smaller than the ALMA $L_{10}^\mathrm{\oiii}/\mathrm{SFR}$ measurement uncertainties. The major cause of this $<0.5$ dex discrepancy is adopting the $Q_\mathrm{\hi}/\mathrm{SFR}-Z$ relation Eq.~(\ref{eq:Q-SFR}), which fails to capture the $Q_\mathrm{\hi}/\mathrm{SFR}$ scatter found in more realistic models such as the FIRE simulations. 
\par

\section{Conclusion and discussion}\label{sec:discussion}
In this work we developed an analytical ionized ISM line emission model that connects [\oiii], [\oii], H$\alpha$, and H$\beta$ line luminosities with the underlying ISM gas properties as well as the incident hydrogen ionizing spectrum. This model treats each \hii\ region as a spherically symmetric sphere with the hydrogen ionizing radiation source located at the centre. It solves for the volumes of the \hii, \oiii, and \oii\ regions assuming ionization--recombination balance among \hi, \hii, \hei, \heii, \oi, \oii, and \oiii. Population abundances among the five energy states of \oiii\ and \oii\ are solved assuming that all ions have achieved a steady state, such that the level population abundances do not vary with time. Compared to publicly available numerical spectral synthesis codes, the strength of this model is its high computational efficiency. For example, our model is 100--1000 times faster than \textsc{Cloudy} in solving the [\oiii] and [\oii] lines without loss of important microphysical processes. The most time consuming part of the model is to solve for the radial profiles of \hi, \hii, \hei, \heii, \oi, \oii, and \oiii\ throughout each \hii\ region. In the publicly available version of this model we provide a lookup table for $V_\mathrm{\oiii}/\tilde{V}_\mathrm{\hii}$ on a fine grid, so that this model can post-process ISM emission line signals for zoom-in galaxies in a few minutes. Our model is therefore suitable for interpreting ISM emission line measurements. The post-processed emission line products are further useful for comparisons among different simulations. Due to the high modeling speed, this framework can also quickly compare variations in the line signals across different environments and for a range of incident radiation spectra. Although in this work we consider uniform \hii\ regions that can each be characterized by a constant gas density, metallicity, and temperature, non-trivial ISM property probability distribution functions can easily be implemented into the model. 
Finally, this semi-analytical model can also be extended to other lines such as [\nii] and [\siii].\par
As an example application, we have employed this model to post process \hii\ region line emission signals for the publicly available FIRE high-$z$ zoom-in simulations. We treat \hii\ regions sourced by stellar particles within each primary galaxy at $z=6$ as individual line emitting regions, and model their [\oiii], [\oii], H$\alpha$, and H$\beta$ line luminosities accounting for variations in the \hii\ region gas density, metallicity, as well as in the shape and strength of the stellar radiation spectrum. We show quantitatively that the [\oiii], H$\beta$, and [\oii] lines trace slightly different stellar populations. 
Among most FIRE galaxies, stellar populations younger than $\sim6$ Myr contribute more than 90\% of the [\oiii] and H$\beta$ line luminosities, while 90\% of the [\oii] signals come from stellar populations younger than $\sim 15$ Myr. We compare the FIRE galaxy-wide [\oiii] 5008\AA, 4364\AA, 88$\mu$m, H$\beta$, and [\oii] 3727,30$\angstrom$ line signals with recent \textit{JWST} and ALMA measurements. We find that simulations and observations are generally in good agreement regarding the line luminosities and luminosity ratios. In summary, the FIRE simulations show that [\oiii] and H$\beta$ lines trace young stellar populations with high hydrogen ionizing photon production rates and relatively hard spectral shapes, while a significant fraction of the [\oii] signal comes from slightly older stellar populations with softer radiation spectra.\par

We tested the common one-zone approximation by fitting one-zone models to our more detailed
line emission calculations from each of the 22 FIRE high-$z$ galaxies. Our post-processed line emission calculations account for spatial variations in the ISM gas density, metallicity, and stellar populations across each galaxy. We then extract ISM parameters from one-zone fits to our FIRE line luminosity models, and test how well the recovered parameters compare with the true simulated values. We consider three scenarios: multiple [\oiii] and [\oii] optical lines are detected by \textsc{JWST}; multiple [\oiii] sub-millimeter lines are observed by ALMA; and a case where only the [\oiii] 88\,$\mu$m line is measured by ALMA. In each case we show quantitatively that the ISM parameters constrained by the one-zone model are close to the simulated ISM properties compared to current measurement uncertainties, thus validating the one-zone assumption. Moreover, we specify the physical meaning of one-zone parameters. Specifically, the metallicities constrained by [\oiii] and [\oii] lines correspond to the \oiii\ and \oii\ ionic abundances $Z_\mathrm{R3}$ and $Z_\mathrm{R2}$, as defined in Eqs.~(\ref{eq:Z_R3})--(\ref{eq:Z_R2}). The combined oxygen abundance corresponds to the $Q_\mathrm{\hi}$-weighted metallicity averaged over the \hii\ regions across a galaxy, $Z_\mathrm{Q}$. Although the values of $Z_\mathrm{Q}$, $Z_\mathrm{R3}$, and $Z_\mathrm{R2}$ depend on the stellar population synthesis model and the approach for estimating ISM properties, we show that the gas particle mass-weighted metallicities of \textsc{FIRE} galaxies $Z_m$ are close to $Z_\mathrm{R3}$ and only slightly lower than $Z_\mathrm{Q}$. This result provides evidence that the observed metallicities can be well-compared with simulated mass-weighted metallicities, $Z_m$. The \hii\ region densities determined from one-zone model fits to  $L_{10}^\mathrm{\oiii}/L_{21}^\mathrm{\oiii}$ are close to the true [\oiii] luminosity-weighted densities in the simulations, and are much higher than the densities averaged over all gas particles. It is therefore difficult to cross-check the gas densities in simulations and observations without ISM emission line post-processing efforts. We also consider one-zone model fits to cases where only [\oiii] 
88\,$\mu$m and SFR measurements are available. In this case, we find that scatter in the $L_{10}^\mathrm{\oiii}/\mathrm{SFR}$ 
ratio is important. Nevertheless, one-zone inferences in the $n_\mathrm{H}-Z$ parameter plane are accurate to better than 0.5\,dex, which is smaller than current ALMA $L_{10}^\mathrm{\oiii}/\mathrm{SFR}$ measurement uncertainties. 
Therefore, we find that one-zone ISM parameter inferences are generally adequate
for [\oiii], [\oii] and hydrogen Balmer lines, provided that the physical meanings of the inferred parameters are interpreted carefully. \par

Although our line emission models agree well with
current [\oiii], [\oii], and H$\beta$ line luminosity measurements, there are still some caveats regarding our modeling. First, we estimate the \hii\ region environmental properties by averaging over the nearest 32 gas particles from each stellar particle. This partly captures inhomogeneities in the relevant ISM properties, but  it still ignores variations across each (sub-grid) \hii\ region. Also some of our results may be influenced by the resolution of the FIRE simulations and the precise choice of averaging over 32 surrounding gas particles. Further, we assume that all of the ionizing photons from each stellar particle are absorbed locally and ignore non-local effects. However, resolution limitations are inherent to simulations so future developments may focus on hybrid methods combining sub-grid and non-local calculations.
We also use single-star population synthesis models from FSPS to assign SEDs to each stellar particle. 
In a more realistic picture, binary stars neglected in this model can produce hard SEDs in older stellar populations
due to mass transfer and mergers \citep[e.g.][]{2016MNRAS.459.3614M}, which might
impact our [\oiii] and [\oii] line luminosity predictions and especially their dependence on the age of the stellar particles in the simulation. The \oii\ versus \oiii\ region temperature model we adopt in this work is calibrated to local measurements. Although direct $T_4^\mathrm{\oii}$ measurements at high redshifts, e.g. $z > 6$, are currently impractical, it will be helpful to better understand this local $T_4^\mathrm{\oii}(T_4^\mathrm{\oiii})$ correlation and predict its potential redshift evolution. Finally, our ISM line emission model lacks a treatment of UV photon absorption by dust and line attenuation from dust. The modeling products are therefore likely most applicable to galaxies of sub-Solar metallicity and should also be compared to 
dust-corrected
line luminosity measurements.\par

In any case, our methodology and extensions should be useful for a broad range of future investigations. For example, comparisons between simulated spectral lines and observed line profiles should provide tests of the kinematic properties of the simulated galaxies. In addition, the galaxy morphologies presented in Figure~\ref{fig:Flux} can be directly compared with \textit{JWST} NIRSPEC IFU (integral field unit)-type observations. 
Another possible application is to combine our models with cosmological simulations to produce mock line-intensity mapping survey data. The SPHEREx (Spectro-Photometer for the History of the Universe, Epoch of Reionization, and Ices Explorer) satellite is expected to launch soon, and among other things, it will produce line-intensity mapping data cubes including measurements of [\oiii], [\oii], and Balmer lines across a wide range of redshifts \citep{2014arXiv1412.4872D}. We will explore these applications of our modeling in future works.\par

\section{Acknowledgement}
We thank Xiangcheng Ma, Andrew Wetzel, Philip Hopkins, and Josh Borrow for useful discussions. We thank the referee, Yuguang Chen, for comments that significantly improved the presentation of this work. AL acknowledges support through NASA ATP grant 80NSSC20K0497. AJB acknowledges support from NASA under JPL
Contract Task 70-711320, ``Maximizing Science Exploitation of
Simulated Cosmological Survey Data Across Surveys''.

\section*{Data availability}
The ISM emission model introduced in this work is publicly available at \url{https://github.com/Sheng-Qi-Yang/HIILines}. The FIRE simulation high-$z$ suite utilized in this work is publicly available at \url{http://flathub.flatironinstitute.org/fire}. Post-processed FIRE galaxy line emission signals are available from
the corresponding author upon request.



%\bibliographystyle{mnras}

\bibliographystyle{mnras}
\bibliography{OIII}

%%%%%%%%%%%%%%%%%%%%%%%%%%%%%%%%%%%%%%%%%%%%%%%%%%

%%%%%%%%%%%%%%%%% APPENDICES %%%%%%%%%%%%%%%%%%%%%

%%%%%%%%%%%%%%%%%%%%%%%%%%%%%%%%%%%%%%%%%%%%%%%%%%

% Don't change these lines
\label{lastpage}
\end{document}